\documentclass[11pt,fleqn]{article} \usepackage{epsfig,color}
\def\greaterthansquiggle{\raise.3ex\hbox{$>$\kern-.75em\lower1ex
\hbox{$\sim$}}}
\def\lessthansquiggle{\raise.3ex\hbox{$<$\kern-.75em\lower1ex
\hbox{$\sim$}}} \newcommand{\grts}{\greaterthansquiggle}
\newcommand{\lets}{\lessthansquiggle} \newcommand{\la}{\label}
\newcommand{\re}{\ref} \newcommand{\ci}{\cite}
\newcommand{\beqn}{\begin{eqnarray}} \newcommand{\eeqn}{\end{eqnarray}}
\newcommand{\bequ}{\begin{equation}} \newcommand{\eequ}{\end{equation}}
\newcommand{\bsl}{\begin{sloppypar}} \newcommand{\esl}{\end{sloppypar}}
\newcommand{\Cdgz}{\ensuremath{\Delta g^\mathrm{Z}_1}}
\newcommand{\Cdkz}{\ensuremath{\Delta \kappa_\mathrm{Z}}}
\newcommand{\Cdkg}{\ensuremath{\Delta \kappa_{\gamma}}}
\newcommand{\Clg}{\ensuremath{\lambda_{\gamma}}}
\newcommand{\Clz}{\ensuremath{\lambda_{Z}}}
\newcommand{\Cgz}[1]{\ensuremath{g^Z_{#1}}}
\newcommand{\Ckzt}{\ensuremath{\tilde{\kappa}_Z}}
\newcommand{\Clzt}{\ensuremath{\tilde{\lambda}_Z}}

\begin{document}
\definecolor{Black}{named}{Black}
\definecolor{Blue}{named}{Blue}
\definecolor{Red}{named}{Red}
\definecolor{Green}{named}{ForestGreen}
\definecolor{Black}{named}{Black}
\definecolor{Olive}{named}{OliveGreen}
\definecolor{Royal}{named}{RoyalBlue}
\definecolor{Orange}{named}{YellowOrange}
\definecolor{Yellow}{named}{Goldenrod}
\definecolor{Cornblue}{named}{CornflowerBlue} 
\definecolor{Lila}{named}{DarkOrchid} 

\setlength{\unitlength}{1cm}
\null
\hfill LC-TH-2000-055\\
\null
\hfill DESY 00-178\\
\null
\vskip .4cm
\begin{center}
{\Large \bf Physics Opportunities with \\[.4em]
Polarized $e^-$ and $e^+$ Beams at TESLA}\\[.4em]
\vskip 1.5em

{\large
{\sc Gudrid Moortgat-Pick$^{a}$\footnote{e-mail:
    gudrid@mail.desy.de},
Herbert Steiner$^{b}$\footnote{e-mail:
    steiner@lbl.gov}
}}\\[3ex]                     
\end{center}
{\footnotesize \it
$^{a}$ DESY, Deutsches Elektronen--Synchrotron, D--22603 Hamburg, Germany}\\
{\footnotesize \it
$^{b}$ Department of Physics and Lawrence Berkeley National Laboratory, 
University of\\\hspace*{.1cm} California, Berkeley, CA 94720, USA }\\
\vskip .5em
\par
\vskip .4cm

\begin{abstract}
Beam polarization at $e^+$--$e^-$ linear colliders will be a powerful tool for 
high precision analyses.  Often it is assumed that the full information from 
polarization effects is provided by polarization of the electron beam and no 
further information can be obtained by the simultaneous polarization of the 
positrons. In this paper we point out the advantages of polarizing both 
beams, and summarize the polarization-related results of the Higgs, 
Electroweak, QCD, SUSY and Alternative Theories 
working groups of the ECFA/DESY
workshop for a planned linear collider operating in the energy range 
$\sqrt{s}= 500-800$~GeV.

\end{abstract}
\vspace{1em}
\hfill
\newpage
\section{Introduction}
\label{sec:1}
Physics beyond the Standard Model (SM) may well be discovered at 
the run II of 
Tevatron, which starts in March 2001, or at the LHC whose start 
is planned for 
2006. However, it is well known that a Linear Collider (LC) will be 
needed for  precise
measurements and 
for the detailed exploration of possible New Physics (NP).
It will also provide access to large uncharted regions
of the parameter space of Grand Unified Theories (GUT). 
 Furthermore a LC will make possible measurements of the SM with 
unprecedented precision.  The main advantages of an $e^- e^+$ collider are the 
clear signatures that make possible precise measurements of the masses and 
couplings of the interacting particles.  Moreover the chiral character of 
the couplings can be worked out at a LC by using beam polarization. The 
importance of such measurements and the physics accessible with polarized 
electrons has been discussed for example in reference \ci{Snowmass96}.

In the limit of vanishing electron mass
SM processes in the s--channel are initiated by
electrons and positrons polarized in the same direction, i.e. $e^+_Le^-_R$ (LR)
or $e^+_Re^-_L$ (RL), where the first (second) entries 
denote helicities of corresponding particle.
This result follows from the vector nature of
$\gamma$ or $Z$ couplings (helicity--conservation).
For these processes positron polarization provides no fundamentally new 
information, but as we will indicate below it can, nevertheless, have 
important consequences \ci{Baltay}. 
In theories beyond the SM interactions also   
(LL) and (RR) configuations from s-channel contributions are allowed and
the polarization of both beams offers a powerful tool for enhancing rates 
and suppressing SM backgrounds. 

Within the framework of the ECFA/DESY Study for a Linear Collider in the 
TeV range the physics consequences when both the 
electron and positron beams are polarized were explored. 
The results show that there are 
five principal advantages to be gained when both beams are polarized: 
(1) higher effective polarization,  (2) suppression of background (3) 
enhancement of rates (${\cal L}$) 
(4) increased sensitivity to non-standard 
couplings, and (5) improved accuracy in measuring the polarization.  These 
features will be discussed in greater detail in the following sections, 
where the polarization-related results of the individual physics working 
groups are summarized.  
Before doing so, however, we briefly indicate in 
Section~\ref{sec:2} 
the experimental possibilities to produce and measure polarization 
at the LC.  In Section~\ref{sec:3} 
we present the results of the different working 
groups. The advantages that might be realized with positron polarization are 
summarized in tabular form in Section~\ref{sec:4}.
\section{Production, Manipulation and Measurement of Polarized Beams at 
a LC}\label{sec:2}
During its last year of running the SLC at SLAC had an average longitudinal 
electron beam polarization $P_{e^-}=77\%$, and polarimetry was developed that 
allowed the polarization to be determined to better than $0.5\%$  
\ci{Baltay}.  
Polarization reversals could be made as desired on a pulse by pulse basis 
by reversing the handedness of the circular polarization of the laser used 
to photo-produce the electrons.  Polarization rotators had to be used to 
prevent depolarization effects in the damping rings.  Very similar techniques
are applicable to LCs more generally, and $P_{e^-}\ge 80\%$ 
polarized electron beams 
together with polarimetry at the $\le 0.5\%$ level should be possible.

Positron polarization, on the other hand, is intrinsically harder. The 
method of choice was originally suggested by Balakin and Mikhailichenko
\ci{Balakin}, 
who proposed to pair-produce longitudinally polarized positrons and electrons
by circularly polarized photons generated by an electron beam in a helical 
undulator.   At TESLA highly polarized $\sim 20$~MeV photons generated by 
$250$~GeV 
electrons passing through a $\sim 100$ meter-long 
helical undulator could be used 
to produce the polarized positrons in a thin foil.  Positron polarizations 
of $40-45$\% are possible with no loss of luminosity.  Higher polarization 
could be achieved at the price of reduced intensity 
(e.g., $P_{e^+} = 60$\% at $\sim 55\%$ of full intensity \ci{Floettmann}).  
Reversing the sign of the polarization is much more difficult in this case.  
A possible way of making reasonably rapid spin reversals might be to use a 
pulsed magnet to steer the positrons into one (or the other) of two separate 
beam lines, each with its own set of solenoidal spin rotators, prior to 
injecting them into the damping rings \ci{Floettmann}.  
No matter what method is used, 
however, care must be taken not to introduce systematic effects in the 
process of flipping the spin.

Polarimeters based on both Moeller and Compton scattering have been used 
to measure electron polarization.  The highest analyzing power, and thus the 
greatest precision ($<0.5\%$), has been attained by detecting the recoil 
electrons when circularly polarized laser photons Compton scatter from 
longitudinally polarized electrons.  
Detection of the scattered photons has 
also been used, but the analyzing power is much smaller and the precision is 
not as high.  Polarimetry using Moeller scattering has the lowest analyzing 
power but it has been used to make measurements at the few percent level.  
Compton scattering is the same for electrons and positrons, so that the same 
polarimeters can be used for both. Detailed design
characteristics of a high-precision polarimeter for use at TESLA are described
in reference \ci{Schuler}.

When both beams are polarized a scheme developed by Blondel obviates the 
need for precision polarimetry.  In this case measurement of the cross 
sections of all spin combinations ((RR), (RL), (LR), and (LL)) can be used 
to determine the effective polarization.  Blondel Scheme 
has the additional advantage 
that the polarization measured in this way is the luminosity-weighted value 
at the interaction point, rather than the value at the location of the 
polarimeter.  With this method it is important that any differences in 
polarization between left- and right-handed electrons (or positrons) be 
carefully measured.  To this end polarimetry will still be needed. 

\section{Results of Polarization Studies for a LC}
\label{sec:3}
Five working groups -- Higgs, Electroweak Theory, QCD, Supersymmetry (SUSY) 
and Alternative Theories -- worked out the physics potential of
a planned linear collider with a first phase of
$\sqrt{s}=500-800$~GeV, which complements and extends the capabilities
of the LHC. In particular the LC provides a well--defined initial state  
and allows the exploitation
of the effects of polarizing the incoming beams.
\subsection{Higgs Working Group}
\label{sec:21}
If a Higgs particle exists in Nature, the accurate study of
its production and decay properties 
can be performed in the 
clean environment of $e^+ e^-$ linear colliders
in order to establish experimentally the Higgs mechanism
as the mechanism of electroweak symmetry breaking 
\cite{DESY-Reports}. The
study of Higgs particles will therefore represent a central
theme of the TESLA physics programme.  
\subsubsection{Separation of production processes}\la{sec:211}
Higgs production at a LC occurs mainly via $WW$ fusion, 
$e^+e^-\to H \nu \bar{\nu}$, and Higgsstrahlung $e^+ e^-\to HZ$. 
Polarizing both beams enhances the signal and suppresses background.
In Table~\ref{tab_higgs1} the scaling factors, i.e.\ ratios of
polarized and unpolarized cross section, are compared for two cases 
 (1) $P_{e^-}=\pm 80\%$, which will be henceforth
denoted by $(80,0)$, 
and (2) $P_{e^-}=\pm 80\%$, $P_{e^+}=\mp 60\%$ denoted throughout the paper
as $(80, 60)$ \ci{Desch_obernai}. 
Here, and in all the subsequent sections of this paper, we will use the 
convention that, if the sign is explicitly given, $+$ $(-)$ polarization
corresponds to R (L) chirality with helicity $\lambda=+\frac{1}{2}$ 
($\lambda=-\frac{1}{2}$) for both electrons and positrons.

If a light Higgs with $m_H\le 130$~GeV is assumed, which is the prefered 
range both
by fits of precision observables in the SM \ci{Delphi} and also 
by predictions of SUSY theories (see e.g. \ci{Weiglein1}), 
Higgsstrahlung 
dominates for $\sqrt{s}\lets 500$~GeV and $WW$ fusion for 
$\sqrt{s}\grts 500$~GeV.
At a LC with $\sqrt{s}=500$~GeV and unpolarized beams the two processes have
comparable cross sections. 
Beam polarization can be used to enhance the WW--Fusion signal 
with respect to the HZ contribution by a factor 1.6 (1.7) if electron (electron
and positron) beam polarization is available (see Table ~\ref{tab_higgs1}).
Further, variation of the relative amounts of
Higgs-strahlung and $WW$ fusion makes it possible to keep
the systematics arising from
the  contributions to the fitted spectrum for these two processes 
smaller than the statistical accuracy.

\subsubsection{Suppression of background}\la{sec:212}
Beam polarization is very effective for background suppression. With 
right-handed electron polarization WW and single Z background can be 
strongly suppressed. The latter is only important at high s.  With 
simultaneous polarization of $e^-$ and $e^+$ 
one gains another factor of about 2 in 
background suppression.  Background from $ZZ$ is slightly suppressed with 
(R0) but not with (RL). To separate the Higgsstrahlung process from $WW$ 
fusion 
right-polarized electron configurations are very suitable. Simultaneous 
left-handed polarized positrons suppress $WW$ fusion and $WW$ background, but 
enhance the $ZZ$ background. Even in the case when the signal to background 
ratio, $S/B$, improves only slightly by simultaneous polarization of the 
$e^-$ and $e^+$ beams with (80,60) one gets an improvement in $S/\sqrt{B}$ 
of about 20\% in  
(Table~\ref{tab_higgs1}). 
Polarization of the positron beams is a powerful tool
to 
suppress the single W  background, $e^+e^-\to W^-e^+\nu$ 
and $e^+e^- \to W^+e^-\bar{\nu}$. 
\subsubsection{Determination of general $ZZ\Phi$ and $Z\gamma\Phi$
couplings}
\la{sec:213}
In an effective Lagrangian approach the general coupling between
Z--, Vector-- and Higgsboson can be written \ci{Kniehl_desy00}:
\beqn
{\cal L}&=&(1+a_Z)\frac{g_Z m_Z}{2} H Z_{\mu} Z^{\mu} +\frac{g_Z}{m_Z}
\sum_{V=Z,\gamma} \big[ b_V H Z_{\mu\nu}V^{\mu\nu}\nonumber\\
&& +c_V(\partial_{\mu} H Z_{\nu}-\partial_{\nu} H Z_{\mu}) V^{\mu\nu}
+\tilde{b}_V H Z_{\mu\nu} \tilde{V}^{\mu\nu}\big],\la{eq_higgs1}
\eeqn
with $V_{\mu\nu}=\partial_{\mu} V_{\nu}-\partial_{\nu} V_{\mu}$,
$\tilde{V}_{\mu\nu}=\epsilon_{\mu\nu\alpha\beta} V^{\alpha \beta}$.

Using, for example, the optimal--observable method \ci{Atwood}
it is possible at a LC
to determine the seven complex 
Higgs couplings with high accuracy: the CP--even
$a_Z$, $b_Z$, $c_Z$ and $b_{\gamma}$, $c_{\gamma}$ and the CP--odd
$\tilde{b}_Z$ and $\tilde{b}_{\gamma}$. 
Simultaneous beam polarization considerably improves the accuracy. 
In \ci{Kniehl_desy00} a study was made for $\sqrt{s}=500$~GeV and 
${\cal L}=300$~fb$^{-1}$. It compares the so--called optimal errors when
tagging 
efficiences $\epsilon_{\tau}=50\%$, $\epsilon_b=60\%$ and beam polarization
and $(\pm 80,\mp 60)$ are assumed.
It shows that the $ZZ\Phi$ coupling is rather well
constrained. However, to fix the $Z\gamma \Phi$ coupling  
beam polarization is essential.
In Table~\ref{tab_higgs2} the optimal errors are listed for the three cases
(1) unpolarized beams, (2) $(\pm 80,0)$ and (3) $(\pm 80,\mp 60)$. 
For this comparison we list the values for
$\epsilon_{\tau}=0\%=\epsilon_{b}$. 
It is obvious that beam polarization is
a decisive tool to fix the general Higgs coupling.
Simultaneous beam polarization
of $e^-$ and $e^+$ beams results in an further reduction 
of 20\%--30\% in the optimal errors
compared to the case $(\pm 80,0)$.

\subsection{Electroweak Working Group}
\label{sec:22}
At a LC it is possible to test the SM with unprecedented accuracy.
At high $\sqrt{s}$ studies determining the triple gauge
couplings \ci{Moenig,Menges} and at low $\sqrt{s}$ 
an order--of--magnitude improvement in the accuracy of
the determination of
$\sin^2\Theta_{eff}^l$ at $\sqrt{s}=m_Z$ may well be possible
\ci{Moenig,Weiglein2}.
These measurements are based on 
the use of polarized beams. To achieve the high precision it will be necessary
to determine the polarization of both beams with very high accuracy.
The Blondel Scheme \ci{Moenig,Blondel} coupled with Compton polarimetry
offers a promising method to realize these goals.
\subsubsection{High $\sqrt{s}$}
The production $e^+ e^-\to W^+ W^-$ occurs 
in lowest order via $\gamma$--, $Z$-- and
$\nu_e$--exchange. In order to test the SM with high precision one can 
carefully study triple gauge boson couplings, which are generally
parametrized in an effective Lagrangian by the C--, P--conserving couplings
$g_1^V$, $\kappa_V$, $\lambda_V$ and the C-- or P-- violating couplings
$g_4^V$, $g_5^V$, $\tilde{\kappa}_V$ and $\tilde{\lambda}_V$ with 
$V=\gamma, Z$ \ci{Hagiwara2}.  In the SM at tree level the couplings
have to be $g_1^V=1=\kappa_V$, while all other are identical to zero.

These couplings can be determined by measuring the angular
distribution and polarization of the $W^{\pm}$'s.
Simultaneously fitting of all couplings results in a strong 
correlation between the $\gamma-$ and $Z-$couplings whereas 
polarized beams are well 
suited to separate these couplings \ci{Menges}.

If the polarization is only known up to 
$\Delta P_{e^-}\sim 1\%$ the statistical error in the gauge 
couplings would be smaller than the error due to 
the uncertainty in determining the beam polarization.
With Compton polarimetry $\Delta P_{e^-}< 0.5\%$ should be possible.
However, to reduce 
the errors further it is necessary to make $\Delta P_{e^-}\ll 1\%$.
This can be done by different schemes, e.g. by measuring the polarization via
the $A_{LR}$ in the forward peak which is dominated by the $\tilde{\nu}$ 
exchange in the $t$--channel and has to be one in this region,
independent of anomalous gauge couplings.
Another possibility is to use simultaneously 
polarized $e^+$ and $e^-$ beams and thereby to reduce the error
in a polarization measurement by a factor three 
if (80,60) is used \ci{Menges}.
The polarization can also be measured by using the Blondel Scheme, 
i.e.\ measuring the configurations (LR), (RL), (LL), and (RR) with
high accuracy. A further advantage of using polarized $e^-$ and $e^+$ beams
is that one could gain about a factor 2 in running time by using the
optimal spin combination.
 
It is well known that TESLA with its high luminosity is a very
promising device to measure these couplings with high precision 
\ci{Moenig}. At $\sqrt{s}=500$~GeV and with $|P_{e^-}|=80\%$ statistical
errors of about $\Delta\kappa_{\gamma}\le 3\times 10^{-4}$ and
$\Delta\lambda_{\gamma}\le 5\times 10^{-4}$ can be reached.
Moreover, using simultaneous beam polarization $(80,60)$ 
the errors can be further reduced by up to a factor 1.8 
compared to the case with $(80,0)$.
\ci{Menges} (see Table~\ref{tab_ew1}).

\subsubsection{GigaZ}
The option
GigaZ refers to running TESLA at the Z resonance with to about $10^9$ Z events.
Beam polarization of both $e^-$ and $e^+$ at GigaZ would make possible the most
sensitive test of the SM ever made.
The potential of this high--precision measurement for testing higher order
quantum effects in the electroweak SM and its supersymmetric extensions
has been studied in \ci{Weiglein2}.
Polarizing both beams has the double
advantage of increasing the effective polarization and significantly reducing
the polarization error. 

In the SM the left--right asymmetry $A_{LR}$ depends only on the effective
leptonic weak mixing angle:
\bequ
A_{LR}=\frac{2(1-4 \sin^2\Theta^l_{eff})}{1+(1-4 \sin^2\Theta^l_{eff})^2}.
\la{eq_ew1}
\eequ
The value of the weak mixing angle listed in the 2001 PDG Tables is
$\sin^2\theta_{eff}=0.23147(16)$ which corresponds to $A_{LR}\sim 0.15$.
In this case the error on $\sin^2\theta_{eff}$ is about a factor of 8 smaller
than the error on $A_{LR}$, i.e.
\bequ
\delta(\sin^2\theta_{eff})\sim \delta(A_{LR})/8.
\label{eq_ew2aab}
\eequ
The error on $A_{LR}$ can be written
\bequ
(\delta(A_{LR}))^2=(\delta(A_{LR}(pol)))^2+(\delta(A_{LR}(stat)))^2
\label{eq_ew2aaaaa}
\eequ
with
\bequ
\delta(A_{LR}(pol))=(A_{LR}/P_{eff})\delta(P_{eff}),
\label{eq_ew2aaaab}
\eequ
where the effective polarization is given by
\bequ
P_{eff}=(P_{e^-}-P_{e^+})/(1-P_{e^-} P_{e^+}).
\label{eq_ew2aaa}
\eequ
From (\ref{eq_ew2aaa}) and (\ref{eq_ew2aaaaa}) we see that with 
$(\pm 80,\mp 40)$, the effective polarization
is increased from 80\% to 91\%, and even more important, the polarization
uncertainty, $\delta(P_{eff})$ is reduced by a factor of 2 compared to the 
value obtained when only the electron beam is polarized. The values of both
$P_{eff}$ and $\delta(P_{eff})$ continue to improve as the positron
polarization increases.

The statistical power of the data sample can be fully exploited only when
$\delta(A_{LR}(pol))<\delta(A_{LR}(stat))$.
For $10^8-10^9$ Z's this occurs when
$\delta(P_{eff})<0.1\%$. In this limit 
$\delta(\sin^2\theta_{eff})\sim 10^{-5}$, which is an order--of--magnitude
smaller than the present value of this error. Thus it will be crucial to
minimize the error in the determination of the polarization.
Improvements in Compton polarimetry might reduce the present error of 0.5\%
by a factor of two, but achieving 0.1\% may be difficult. 
The desired precision should, nevertheless, be attainable with
the Blondel Scheme, where it is not necessary to 
know the beam  polarization with such extreme accuracy, since
$A_{LR}$ can be directly expressed via cross sections for
producing Z's with longitudinally polarized beams:
\beqn 
\mbox{\hspace{-1cm}}
\sigma&=&\sigma_{unpol}[1-P_{e^-} P_{e^+}+A_{LR}(P_{e^+}-P_{e^-})], 
\la{eq_ew2a}\\
\mbox{\hspace{-1cm}}
A_{LR} &=& \sqrt{\frac{(\sigma^{RR}+\sigma^{RL}-\sigma^{LR}-\sigma^{LL})
(-\sigma^{RR}+\sigma^{RL}-\sigma^{LR}+\sigma^{LL})}
{(\sigma^{RR}+\sigma^{RL}+\sigma^{LR}+\sigma^{LL})(-\sigma^{RR}+\sigma^{RL}
+\sigma^{LR}-\sigma^{LL})}}.
\la{eq_ew2b}
\eeqn
In this formula the absolute polarisation values of the left-
and the right-handed states are assumed to be the same. 
Corrections have to be determined experimentally by means of polarimetry
techniques; however, only relative measurements are needed,
so that the absolute calibration of the polarimeter cancels \ci{Moenig}.
Polarimetry is also needed to
track the time dependence of the polarisation which could affect 
the attainable precision in this scheme.

As can be seen from (\ref{eq_ew2b}) the Blondel scheme also requires some
luminosity for the less favoured combinations (LL, RR). However only about
10\% of running time will be needed for these combinations
to reach the desired accuracy for these high precision measurements.
Fig.~\ref{fig_ew1} 
shows the statistical error on $A_{LR}$ as a function of the positron
polarisation for $P_{e^-}=80\%$. 
Already with 20\% positron polarisation the goal
of $\delta \sin^2\theta_{eff} \sim 10^{-5}$ can be reached.
 
As an example of the potential of the GigaZ $sin^2\theta_{eff}$ 
measurement Fig.~\ref{fig_ew2}
compares the present experimental accuracy on $sin^2\theta_{eff}$ and $M_W$
from LEP/SLD/Tevatron and the prospective
accuracy from the LHC and from a LC without GigaZ option
with the predictions of the SM and the MSSM. With GigaZ a very sensitive
test of the theory will be possible.

\subsection{QCD Working Group}
\label{sec:23}
Strong--interaction measurements at TESLA will form an important component of 
the 
physics programme. The LC offers the possibility of testing QCD at high energy 
scales in the experimentally clean, theoretically tractable $e^+ e^-$ 
environment. 
For the  LC $\gamma \gamma$, $\gamma e^-$ and $e^- e^-$ collisions are
conceivable, and these could be used to
study structure functions of photons.
Since these modes require a future upgrade we confine ourselves here to only
the $e^+ e^-$ mode.

The top quark is by far the heaviest fermion observed, yet all the experimental
results tell us that it behaves exactly as would be expected for a third 
generation quark.
Its large mass, which is close to the scale of electroweak symmetry breaking, 
renders
the top quark a unique object for studying the fundamental interactions in the
attometer regime. It is likely to play a key role in pinning down the origin 
of 
electroweak symmetry breaking.
High precision measurements of the properties and the interaction
of top quarks will therefore be an essential part of the LC research program.

\subsubsection{Polarization effects in production of light quarks}
The scaling factors, $\kappa$, for the production of the light quarks,
$e^+ e^-\to q \bar{q}$ with $q=u$, $d$, $c$, $s$, $b$, are given for
$\sqrt{s}=500$~GeV by \ci{Brandenburg}:
\bequ
\kappa=\frac{\sigma^{pol}}{\sigma^{unpol}}=
[1-P_{e^-} P_{e^+}][1- 0.46 P_{eff}],
\la{eq_qcd1}
\eequ
where the factor 0.46 is valid for the sum of light quarks.
In Table~\ref{tab_qcd1} scaling factors for different configurations are
listed and one can see that for left--handed electron polarization 
one gains about 50\%  in rate if one simultaneously uses right--handed 
positrons (80,60). 
Conversely, for right--handed electrons the scaling factor increases
by about 30\% when positron have left--handed polarization.

The main background comes from $e^+ e^-\to W^+ W^-$. The scaling
factors for this reaction are shown in Table~\ref{tab_higgs1}. 
The spin configuration (LR) improves the ratio for quark production. 
However, if right--handed electrons
are used, the $WW$ background is strongly suppressed, whereas for the
configuration (RL) quark production is suppressed by a factor 0.87.
In this case 
$S/B$ improves by nearly a factor of 3 and
$S/\sqrt{B}$ by a factor of 2.

In \ci{Kuehn} polarization effects were studied at the top threshold.
To determine the top vector couplings, $v_t$ one has to measure the 
left--right asymmetry $A_{LR}$
with high accuracy. With an integrated luminosity of ${\cal L}=300$~fb$^{-1}$ 
precisions in $A_{LR}$ and $v_t$ of about $0.4\%$ and $1\%$ respectively can 
be 
achieved at the LC. The gain in using
simultaneously polarized $e^-$ and $e^+$ 
beams $(80,60)$ is given by the higher 
effective 
polarization of $P_{eff}=0.946$ compared to the case for only polarized 
electrons. This enhancing of effective polarization 
is also significant when measuring the decays of 
polarized top quarks at rest, where the lepton angular distribution of 
$t\to b W^+ (\to e^+ \nu)$ is sensitive to the 
Lorentz structure of the charged currents. 
The angular distribution of $t\to H^+ b$ can be used to directly determine
$\tan\beta$.
With ${\cal L}=300$~fb$^{-1}$ the coefficients of the angular dependent terms 
can be determined with accuracies  of about 1--2\%.

\subsubsection{Polarization effects in limits for top FCN couplings}
Recently a study concerning polarization effects of simultaneously 
polarized $e^-$ and $e^+$ beams in measuring
top flavour changing neutral (FCN) couplings 
has appeared \ci{Aguilar}. In Table~\ref{tab_qcd2} the  limits on
top FCN decay branching ratios obtained from single top production,
$e^+ e^-\to t \bar{q}$, are listed. The study was made at $\sqrt{s}=500$~GeV 
with ${\cal L}=300$~fb$^{-1}$ and $\sqrt{s}=800$~GeV with  
${\cal L}=500$~fb$^{-1}$. With $e^-$ and $e^+$ polarization (80, 45), 
limits are improved by about a factor 2.5 compared to no polarization, 
wheras in each case the 
positron polarization improves the limits obtained with only electron
polarization by 30\%--40\%, Table~\ref{tab_qcd2}. Analogous results can be
obtained at $\sqrt{s}=800$~GeV. 
These improvements, due to the use of polarized $e^+$ and $e^-$ beams
correspond to an increase in rate of a factor of 6--7. 
Comparison with the limits for FNC couplings
from LHC show that the LC measurements are  
complementary, perhaps even decisive e.g. for the 
$\gamma t u$--coupling \ci{Aguilar}. 

The dominant background for the $t \bar{q}$ signal is given by $W^+ q
\bar{q}'$ production followed by
$W^+$ decays into lepton pairs. With 
$(80,0)$
background decreases by a factor of 5 while keeping 90\% of the signal.
With $(80,45)$ the background is reduced
by a factor of 8 and the signal is increased by 20\% compared to the
values for unpolarized beams \ci{Aguilar}. 
\subsubsection{Polarized structure functions}
Up to now nothing is known about polarized structure functions (PSF) 
of photons but a LC, especially when operated in the $\gamma \gamma$ or
$e^- \gamma$ mode, is well suited to make such studies.
For TESLA the $\gamma \gamma$ option has been discussed
as a possible upgrade, but it is already possible to get information about
PSF even in the normal $e^+ e^- $ mode if one uses highly polarized $e^+$ and 
$e^-$ beams.

In \ci{Stratmann} studies of PSF were made in 
$e^+ e^-\to \gamma \gamma + e^+ e^-\to \mbox{Di-jets}+e^+e^-$, 
using polarized beams.
It was found that the structure functions could be extracted from 
measurements of 
the asymmetry of the di--jets, $A^{2-\rm{jet}}$. 
Both of the $\gamma$'s have to be polarized, and because
depolarization tends to be large at the $e \gamma$ vertex one
needs highly polarized $e^-$ and $e^+$ beams. In Fig.~\ref{fig_qcd1} 
we show the results of the
study for (70,70) and $\sqrt{s}=500$~GeV.
Unfortunately the di--jet asymmetries, $A^{2-\rm{jet}}$, are only about $1\%$,
and with a data sample of ${\cal L}_{ee}=10$fb$^{-1}$ the errors are 
$0.5\%$. 
With ${\cal L}_{ee}=100$fb$^{-1}$, which should be reachable at
TESLA the errors would be reduced by a factor of three.
Further improvements are possible by optimizing cuts on the data.
Thus, if systematic effects can be kept under control,
at least some initial information about PSF 
may be obtainable at a LC in the $e^+ e^-$ mode. 
For this purpose it may be necessary to operate with the highest possible
positron polarization even at the expense of reduced beam intensity 
\ci{DeRoeck}.

\subsection{Alternative Theories Working Group}
\label{sec:25}
Beam polarization is a helpful tool to enlarge the discovery reach of $Z'$, 
$W'$ and to look for the effects of extra dimensions. Also discrimination 
between different contact interactions should be simplified with the help 
of beam 
polarization.
\subsubsection{Polarization effects in discovery reach for $Z'$ and $W'$}
Studies have been made for direct and indirect evidence for production of Z' bosons 
at a linear collider having $\sqrt{s}=500$~GeV \ci{Riemann}.
Beam polarization enlarges 
the discovery reach but the predicted effects are strongly model dependent. 
In Table~\ref{tab_alt1} the lower bound for $m_{Z'}$ at a LC with
${\cal L}=500$~fb$^{-1}$ are presented for two different models,
one a superstring--inspired E$_6$ model and the other a LR model with
an additional neutral gauge boson $Z'$.
It can be seen  from Table~\ref{tab_alt1} that
with $(80,60)$ the discovery reach is increased by 
$10\%$--$20\%$ compared to the case when $(80,0)$. 
The effects of positron polarization in the 
production of $W'$ bosons are not yet available. In Table~\ref{tab_alt1}
we summarize the results of studies made for
unpolarized beams, and for polarized  electrons with 
$P_{e^-}=80\%$ \ci{Godfrey} for a LC with ${\cal L}=1$ ab$^{-1}$. 
We estimate a roughly 10\% increase in the production of 
left--chirality--prefering 
$W'$ bosons when 
$(80,60)$ as compared to when only $(80,0)$,  
since the effective polarization is increased from $80\%$ to $95\%$.

In search for right--handed neutral gauge bosons polarization will not only
to increase the rate, but even more important it can be used to make major 
reductions in backgrounds caused by standard left--handed interactions. For 
example, with $P_{eff}=95\%$ the left--handed background is reducced by a 
factor of twenty compared to the case of unpolarized beams.

\subsubsection{Polarization effects in contact interactions}
Beam polarization is a useful tool to look for the existence of contact 
interactions
and to distinguish between different models.
Simulation studies are given in \ci{Riemann} for
$\sqrt{s}=500$~GeV and ${\cal L}=1$~fb$^{-1}$.
In Figure~\ref{fig_alt1} the expected sensitivity from
$e^+ e^-\to b \bar{b}$ is listed for different models of couplings.
Systematical errors are included. One sees 
that using $(80,40)$ 
instead of only $(80,0)$
could enlarge the discovery reach for the scale $\Lambda$ by up to $40\%$ 
for RR or RL interactions. 
\subsubsection{Polarization effects in discovery reach for extra dimensions}
In the search for extra dimensions,
$e^+ e^-\to \gamma G$, beam polarization enlarges the discovery reach for 
the scale
$M_D$ \ci{Vest}, and is a crucial tool for suppressing
the dominant background $e^+ e^-\to \nu \bar{\nu} \gamma$ \ci{Wilson}.
In Table~\ref{tab_alt2} the lower bounds of $M_D$ are given for various  
numbers of extra dimensions. The study was made
for $\sqrt{s}=800$~GeV and ${\cal L}=1$~ab$^{-1}$.
With simultaneous beam polarization
of $(80,60)$ the reach is enlarged by 
$16\%$ compared to the case 
with only electron polarization. With increasing numbers of
extra dimensions this factor decreases up to $10\%$.
In Figure~\ref{fig_alt2} 
the cross section for $e^+ e^-\to G \gamma$ as a function
of $M_D$ is plotted for different systematic errors \ci{Vest}. 
For example with unpolarized beams in the case of two extra dimensions, 
$\delta=2$,
and $M_D=7.5$~GeV the signal, S, is about three orders of magnitude smaller 
than the background, B \ci{Wilson}.
However, if polarized beams are used the ratio 
$\frac{S}{\sqrt{B}}$ is improved by 2.2 for $(80,0)$ and
by 5.0 for $(80,60)$ as shown in Figure~\ref{fig_alt3}.
This corresponds to an increase in rate by a factor 5 compared to when  
only electrons are polarized, and a factor 25 when both beams are polarized. 
In studies for indirect limits on extra dimensions
the simultaneous polarization of both beams does not play a significant
role. In this case $P_{e^+}$ improves 
the limits by only $8 \%$ \ci{Riemann2}. 
  
\subsection{SUSY Working Group}
\label{sec:24}
In SUSY models all coupling structures consistent with Lorentz invariance
should be considered. Therefore it is possible to get appreciable 
event rates for polarization configurations that are unfavorable for SM 
processes.
Polarization effects can play a crucial role in discovering SUSY and in the
determination of supersymmetric model parameters. 
All numerical values quoted below, if not otherwise stated, 
are given for the LC--reference scenario for low $\tan\beta$
with the SUSY parameters $M_2=152$~GeV, $\mu=316$~GeV, $\tan\beta=3$ and 
$m_0=100$~GeV \ci{Blair}.
\subsubsection{Polarization Effects in the Stop Sector}
In \ci{Kraml} the feasibility of 
determining the stop mixing angle 
in the process $e^+ e^- \to \tilde{t}_1 \tilde{t}_1$ at the LC has been 
investigated.
The study was made at $\sqrt{s}=500$~GeV, ${\cal L}=2\times 500$~fb$^{-1}$
and polarization
 $(80,60)$ for the parameters $m_{\tilde{t}_1}=180$~GeV,
$\cos\Theta_{\tilde{t}_1}=0.57$. The resulting errors are
 $\delta(m_{\tilde{t}_1})=1.1$~GeV and
$\delta(\cos\Theta_{\tilde{t}_1})=0.01$. If only polarized electrons  
were used then these errors
would increase by about 20\% \ci{Kraml}.
\subsubsection{Polarization effects in slepton production}
With beam polarization the nature of the 
SUSY partners of the left-- and right--handed
particles can be identified.
The production of sleptons $e^+ e- \to \tilde{\ell}_L \tilde{\ell}_L$,
$e^+ e- \to \tilde{\ell}_R \tilde{\ell}_R$
proceeds via $\gamma$ and $Z$ exchange in the direct channel and
$\tilde{\chi}^0_i$ exchange in the crossed channels.
The process $e^+ e- \to \tilde{e}_L \tilde{e}_R$ is only possible via the 
crossed channel.
Beam polarization is a valuable tool to separate $\ell_L$ from $\ell_R$ 
\ci{Dima}.
In our reference scenario, as
one can see in Figure~\ref{fig_susy6}a, the production rates 
for $\tilde{e}_L \tilde{e}_L$, $\tilde{e}_L \tilde{e}_R$, 
$\tilde{e}_R \tilde{e}_R$ are very sensitive to the polarization 
configuration. We distinguish between $\tilde{e}^+_R\tilde{e}^-_L$ and
$\tilde{e}^+_L\tilde{e}^-_R$ since their cross sections can be separated 
via the charge and their different 
dependence on beam polarization \ci{Ghodbane}: 
\begin{itemize}
\item 
$\tilde{e}^+_L \tilde{e}^-_L/\tilde{e}^+_R \tilde{e}^-_R/\tilde{e}^+_R
\tilde{e}^-_L/\tilde{e}^+_L\tilde{e}^-_R=23/15/9/1$ when $(-80,0)$;
\item
$\tilde{e}^+_L \tilde{e}^-_L/\tilde{e}^+_R \tilde{e}^-_R/\tilde{e}^+_R
\tilde{e}^-_L/\tilde{e}^+_L\tilde{e}^-_R
=22/11/2/1$ when $(-80,+60)$ i.e.\,  
the rate for $\tilde{e}^+_L \tilde{e}^-_L$ is enhanced
by a factor 1.6,
whereas $\tilde{e}^+_R \tilde{e}^-_R$ changes only slightly and
the mixed selectron pairs are strongly suppressed;
\item 
$\tilde{e}^+_L \tilde{e}^-_L/\tilde{e}^+_R \tilde{e}^-_R/\tilde{e}^+_R
\tilde{e}^-_L/\tilde{e}^+_L\tilde{e}^-_R
=24/33/36/1$ with $(-80,-60)$; i.e.\,
the rate for $\tilde{e}^+_R \tilde{e}^-_L$ is enhanced by a facor 1.6 
whereas all other slectron pairs are suppressed.
\end{itemize}

In MSSM scenarios one generally finds that 
$m_{\tilde{\ell}_R}<m_{\tilde{\ell}_L}$, because
only in extended models can this 
mass hierarchy be weakened \ci{Ramond}. 
Therefore one can distinguish between the left and right
selectron in the pair by measuring the threshold cross sections 
$(\tilde{e}_R, \tilde{e}_R)$, $(\tilde{e}_R, \tilde{e}_L)$ and 
$(\tilde{e}_L, \tilde{e}_L)$. The two--body
kinematics  allows the identification and accurate measurement of the 
masses, branching ratios, couplings and mixing parameters.

Polarisation is indispensable to determine the weak quantum numbers $R$, $L$ 
of the sleptons \ci{Ghodbane}.
Polarization of both beams facilitates the identification of
the superpartners of left and right selectrons
in a straight--forward manner. S--channel processes are suppressed when both 
beams
are left polarized or both beams are right polarized.
In the limit of completely left (or right) polarized
beams slepton production can only occur via the t--channel. Measurement 
of the threshold curves and identification of the sleptons via charge 
provides a simple and unique method to identify
the SUSY partners of left--, right--handed $e^-$ and $e^+$. 
For right polarized electrons we get the ratios of cross sections shown in  
Figure~\ref{fig_susy6}b:
\begin{itemize}
\item
$\tilde{e}^+_L \tilde{e}^-_L/\tilde{e}^+_R \tilde{e}^-_R/\tilde{e}^+_R
\tilde{e}^-_L/\tilde{e}^+_L\tilde{e}^-_R
=6/53/1/9$ for $(+80,0)$,\\[-.7cm]
\item
$\tilde{e}^+_L \tilde{e}^-_L/\tilde{e}^+_R \tilde{e}^-_R/\tilde{e}^+_R
\tilde{e}^-_L/\tilde{e}^+_L\tilde{e}^-_R
=4/52/1/2$ for $(+80,-60)$
i.e.\,  
the rate for $\tilde{e}^+_R \tilde{e}^-_R$ is enhanced
by a factor 1.6;
\\[-.7cm]
\item $\tilde{e}^+_L \tilde{e}^-_L/\tilde{e}^+_R \tilde{e}^-_R/\tilde{e}^+_R
\tilde{e}^-_L/\tilde{e}^+_L\tilde{e}^-_R
=14/56/1/36$ with $(+80,+60)$ i.e.\,
the rate for $\tilde{e}^+_L \tilde{e}^-_R$ is enhanced by a facor 1.6 
whereas all other slectron pairs are suppressed. 
\end{itemize}
Simultaneous polarization of both beams with $(80,60)$
can enhance the signal by a factor 1.6
compared to the case with only electron polarization \ci{Ghodbane}.

In the reaction $e^+ e^-\to \tilde{\mu} \tilde{\mu}$, only the pairs 
$\tilde{\mu}_L \tilde{\mu}_L$, $\tilde{\mu}_R \tilde{\mu}_R$ are produced
because the $t$--channel drops out. The scaling factors  
are about the same as in selectron production.
\subsubsection{Polarization effects in the chargino sector}
\begin{figure}
\begin{minipage}{7cm}
\begin{picture}(5,5)
\put(0,0){\includegraphics{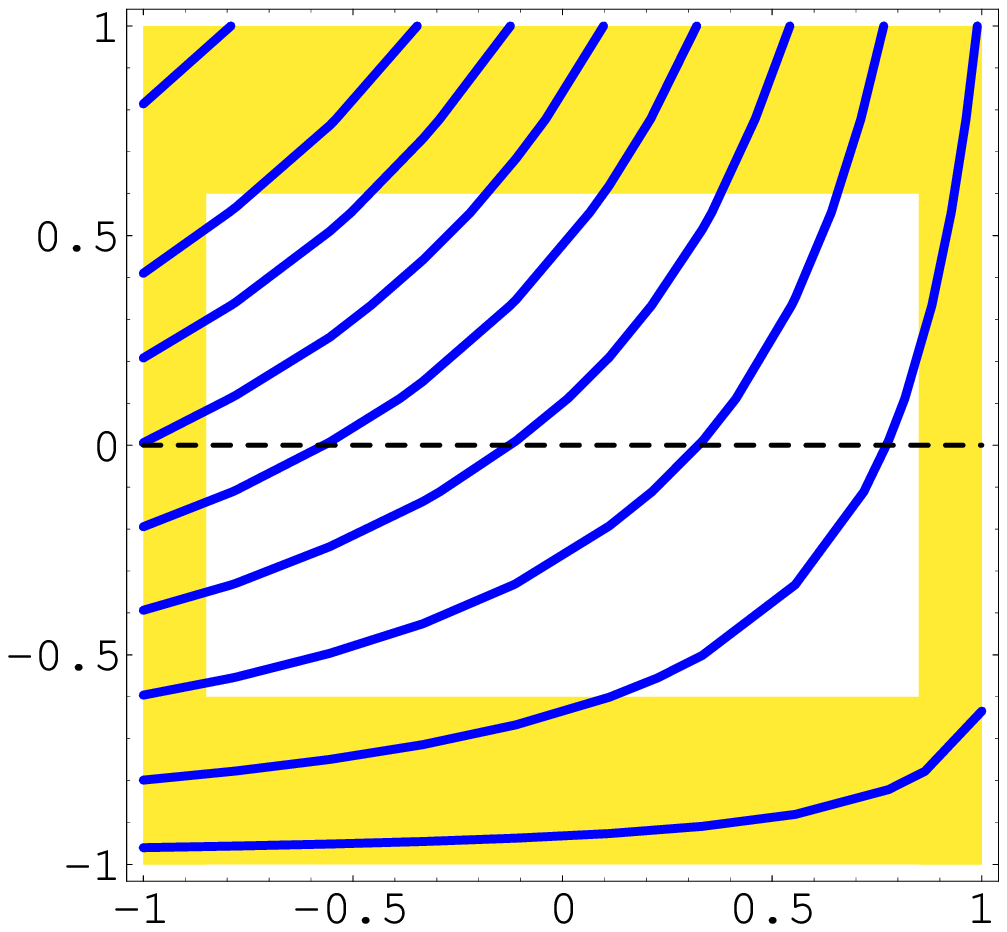}}
\put(1.8,4.8){\small a) $\sigma(\tilde{\chi}^+_1 \tilde{\chi}^-_1)/fb$  }
\put(6,-1.2){ \small $ P_{e^-}$}
\put(0,4.6){\small $ P_{e^+}$}
\put(5.55,-.4){\tiny 10}
\put(4.65,.6){\tiny 50}
\put(3.8,1.2){\tiny 100}
\put(3.2,1.7){\tiny 150}
\put(2.55,2.1){\tiny 200}
\put(2.2,2.55){\tiny 300}
\put(1.9,2.95){\tiny 350}
\put(1.5,3.3){\tiny 400}
\put(1.3,3.7){\tiny 450}
\put(1.05,4.0){\tiny 500}
\end{picture}\par\vspace{1.cm}
\end{minipage}\hfill\hspace{-.6cm}
\begin{minipage}{7cm}
\begin{picture}(5,5)
\put(0,0){\includegraphics{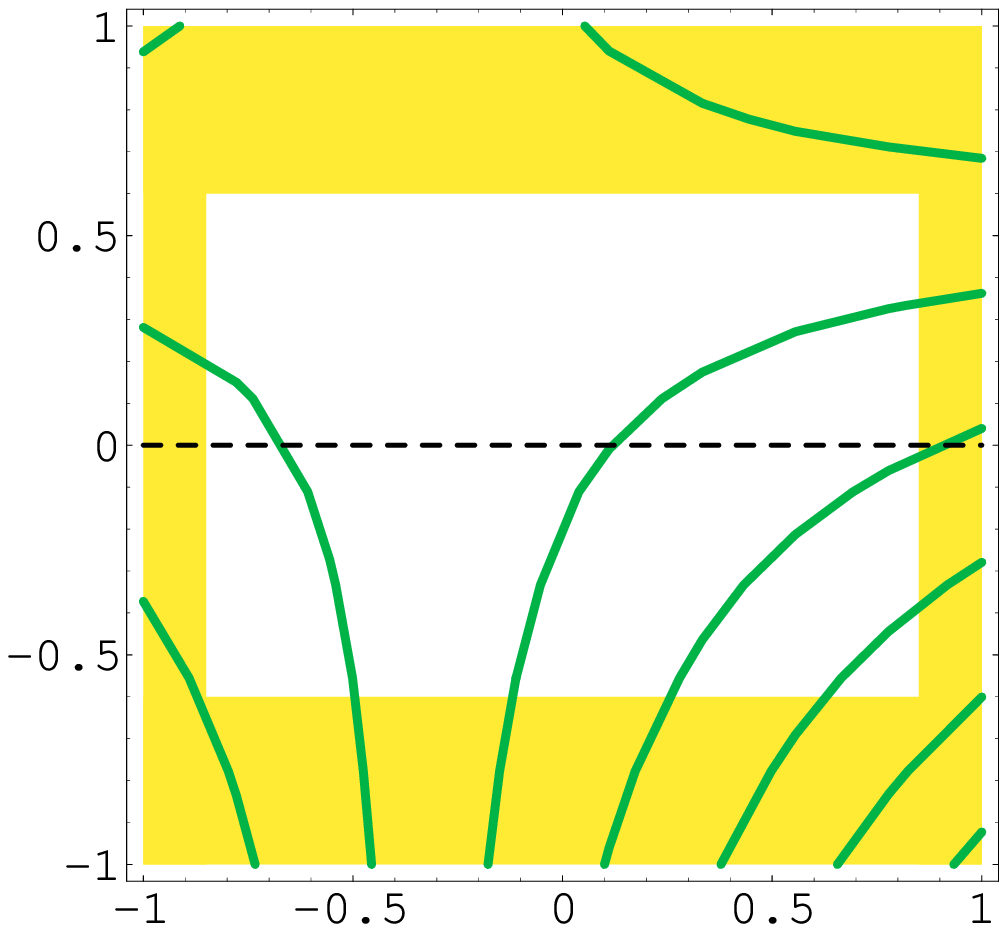}}
\put(1.8,4.8){\small b) $\sigma(\tilde{\chi}^+_1 \tilde{\chi}^-_2)/fb$  }
\put(6.2,-1.2){\small $ P_{e^-}$}
\put(0,4.6){\small $ P_{e^+}$}
\put(1,4.2){\tiny 3}
\put(3.9,3.9){\tiny 2}
\put(1.6,2.3){\tiny 2}
\put(1.3,0.6){\tiny 1}
\put(3.6,2.2){\tiny 3}
\put(4.3,1.1){\tiny 4}
\put(4.8,0.55){\tiny 5}
\put(5.2,-.05){\tiny 6}
\put(5.7,-.4){\tiny 7}
\end{picture}\par\vspace{.6cm}
\hspace{-.5cm}
\end{minipage}
\hspace*{-1cm}\parbox{15cm}{\caption{SUSY -- Chargino production:
Cross sections of $e^+ e^-\to \tilde{\chi}^+_1 \tilde{\chi}^-_1$,
$e^+ e^-\to \tilde{\chi}^+_1 \tilde{\chi}^-_1$ in fb
at $\sqrt{s}=m_{\tilde{\chi}^+_i}+m_{\tilde{\chi}^-_j}+10$~GeV in the 
reference 
scenario for different
electron (positron) beam polarizations,
$P_{e^-}$ ($P_{e^+}$). The white region is accessible for
$|P_{e^-}|\le 85\%$, $|P_{e+}|=60\%$ \ci{Gudi_Char_Neut}.  \label{fig_susy7}}}
\par\vspace{-.4cm}
%
\hspace{-.9cm}
\begin{minipage}{7cm}
\vspace{1.6cm}
\begin{picture}(7,5)
\put(-.1,0){\includegraphics{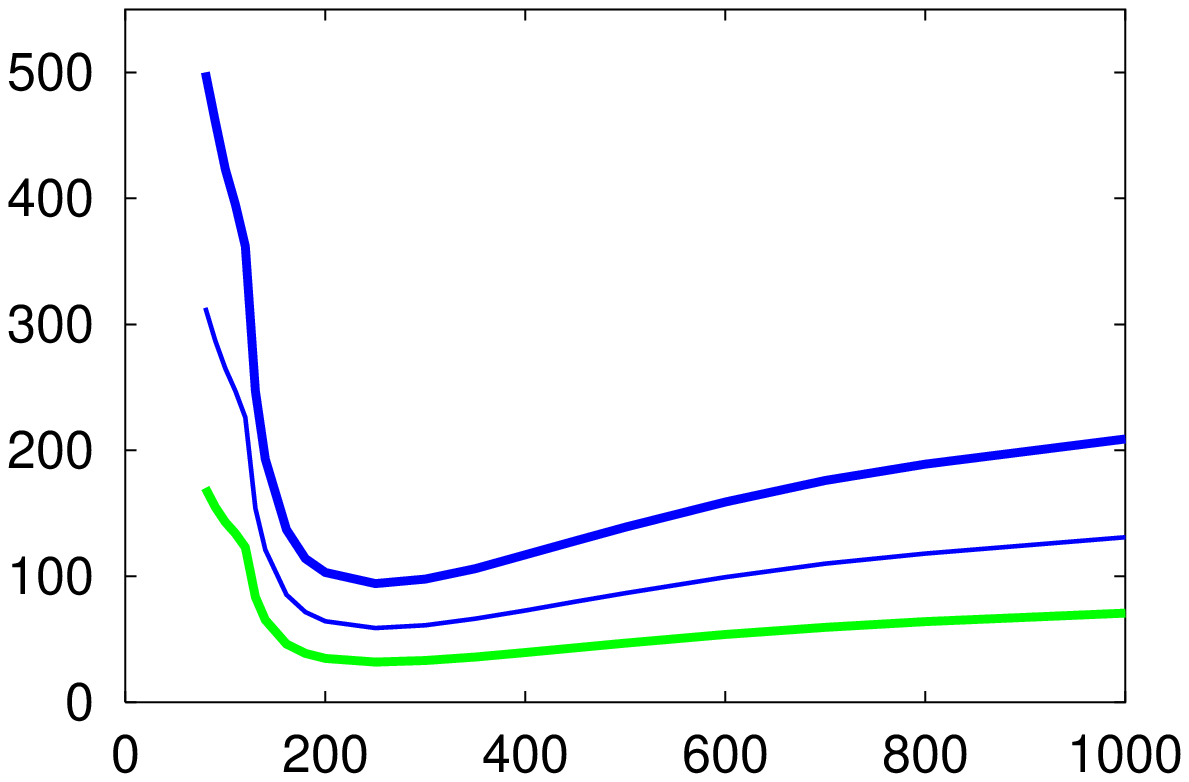}}
\put(2.4,5.1){b) $\sqrt{s}=500$~GeV}
\put(5.8,-.2){\small $ m_{\tilde{\nu}_e}${\small /GeV}}
\put(-.1,5.2){\small $ \sigma_e${\small /fb}}
\put(5.9,1.25){\footnotesize $(00)$}
\put(5.8,1.7){\small (L0)}
\put(5.8,2.6){\small (LR)}
\end{picture}\par\vspace{-.2cm}
\hspace{-.5cm}
\end{minipage}
\hspace*{.1cm}
\begin{minipage}{7cm}
\vspace{1.2cm}
\begin{picture}(7,5)
\put(-.2,0){\includegraphics{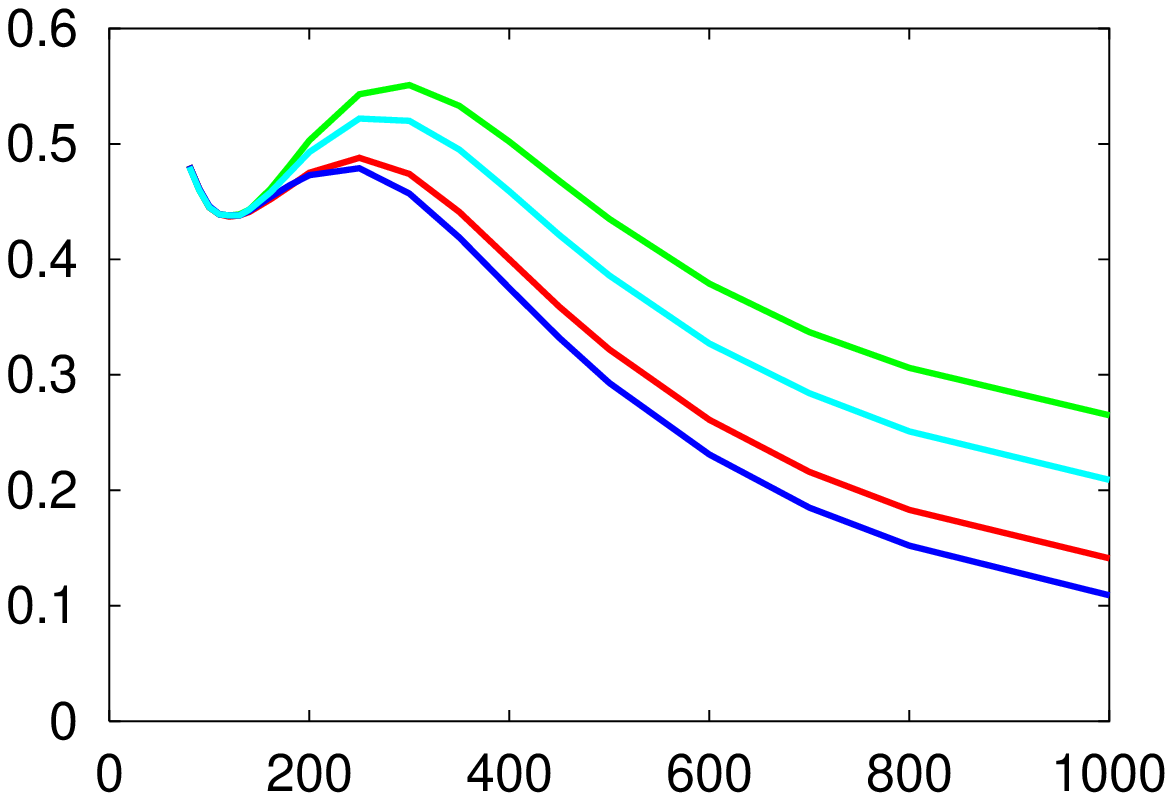}}
\put(2.4,5.1){b) $\sqrt{s}=500$~GeV}
\put(3.2,4.4){\scriptsize $P_-^3=-85\%$, $P_+^3=+60\%$}
\put(5.6,-.2){\small $ m_{\tilde{\nu}_e}${\small /GeV}}
\put(-.1,5.2){\small $ A_{FB}$}
\end{picture}\par\vspace{-.2cm}
\end{minipage}
\hspace*{-1cm}\parbox{15cm}{\caption{ SUSY -- Determination of the 
sneutrino mass 
$m_{\tilde{\nu}_e}$ in
chargino production and decay:
In a) the cross sections $\sigma \times BR$ in [fb] are given for different
 polarization configurations with $|P_{e^-}|=80\%$, $|P_{e^+}|=60\%$
and in
b) the forward--backward asymmetry 
$A_{FB}$ of the decay electron in 
$e^+ e^-\to \tilde{\chi}^+_1\tilde{\chi}^-_1$,
$\tilde{\chi}^+_1\to \tilde{\chi}^0_1 e^- \bar{\nu}$ is
  shown for different masses of sneutrinos $m_{\tilde{\nu}}$ and selectrons
$m_{\tilde{e}_L}=130$~GeV, $m_{\tilde{e}_L}=150$~GeV, 
$m_{\tilde{e}_L}=200$~GeV and 
$m^2_{\tilde{e}_L}=m^2_{\tilde{\nu}}-m_W^2 \cos 2\beta$ 
(from top to bottom line)
\ci{Gudi_Char_Neut}. 
Other SUSY parameters as in the reference scenario. \la{fig_susy8}}}
\end{figure}

As one can see in Fig.~\ref{fig_susy7} beam polarization can
considerably enhance
the cross section for chargino production. For $(80,0)$ the event 
rate is
enhanced by about a factor 1.8. Simultaneous positron polarisation of 
$P_{e^+}=60\%$
leads to an additional increase of a factor 1.6.
This fact, as in the cases before, can not be expressed by the effective 
polarization, because these rates depend explicitly on the polarizations 
of both beams.

In the MSSM the chargino sector depends on the fundamental parameters
$M_2$, $\mu$, $\tan\beta$. Beam polarization is crucial for determining 
these parameters \ci{Choi}. The analysis 
was made for completely longitudinally 
polarized beams and the assumption that the masses of the exchanged
sneutrinos $m_{\tilde{\nu}}$, are known. Using these assumptions 
it has been shown \ci{Choi} 
that these parameters can be determined quite well. 
Since completely polarized beams do not exist, any
efforts to increase the 
effective polarization are important to improve the accuracy of the results, 
which means that positron polarization
is very advantageous. 

In one method to constrain $m_{\tilde{\nu}}$ indirectly 
\ci{Gudi_Char_Neut}, beam
polartization can be used to
help decrease the statistical error, Figure~\ref{fig_susy8}. 
There  it is shown that
the forward--backward--asymmetry of the decay electron in 
$e^+ e^-\to \tilde{\chi}^+_1 \tilde{\chi}^-_1$, 
$\tilde{\chi}^-_1\to \tilde{\chi}^0_1 e^- \bar{\nu}$, is very sensitive to 
$m_{\tilde{\nu}}$. With additional positron beam polarization 
there is a further
increases in the cross section by a factor of about 1.6, so that 
the statistical error in 
$\Delta A_{FB}$ is reduced by 20\%.  

In single chargino production, $e^+ e^-\to \tilde{e} \tilde{\chi}^- \nu_e$,
$e^+ e^-\to \tilde{e} \tilde{\chi}^+ \bar{\nu}_e$ \ci{Baltay}
the preferred beam polarisation configurations 
are (RR) and (LL), which are not 
favored in the SM. The event rates are 
expected to be small so that positron polarization could play a major 
role in the measurement and analysis of this process.

\subsubsection{Polarization effects in the neutralino sector}
Here, too, beam polarization can significantly 
enhance the signal and there\-fore improve the ratios $S/B$ and $S/\sqrt{B}$.
As can be seen in Figure~\re{fig_susy2}a the  
cross section $\sigma(\tilde{\chi}^0_1\tilde{\chi}^0_2)$
can be enhanced by about 60\% for $(-80,+60)$ compared to the case 
$(-80,0)$. For right--polarized
electrons similar results are obtained e.g.\ for
$\sigma(\tilde{\chi}^0_2\tilde{\chi}^0_2)$, Figure~\re{fig_susy2}b. 
For this process an even greater
advantage of polarizing both beams with different signs
is the suppression of the dominant $WW$ background.

Beam polarization may be crucial for the determination  
of the parameters and the couplings 
of the neutralino sector of supersymmetric models.
In lowest order neutralino production occurs via $Z$, $\tilde{e}_L$ and
$\tilde{e}_R$ exchange, so it is sensitive to the chiral couplings and  
the masses of $\tilde{e}_L$, $\tilde{e}_R$. The ordering
of the cross sections for different polarization configurations
depends on the character of the neutralinos:
\begin{itemize}
\item Pure higgsino:\\[-.8cm]
\bequ
\sigma^{LR}>\sigma^{RL}>\sigma^{L0}>\sigma^{00}>\sigma^{R0}>\sigma^{LL}>\sigma^{RR}
\la{eq_susy1a}
\eequ
\item Pure gaugino and $m_{\tilde{e}_L}\gg m_{\tilde{e}_R}$:\\[-.8cm]
\bequ
\sigma^{RL}>\sigma^{R0}>\sigma^{00}>\sigma^{RR}>\sigma^{LL}>\sigma^{L0}>\sigma^{LR}
\la{eq_susy1b}
\eequ
\item Pure gaugino and $m_{\tilde{e}_L}\ll m_{\tilde{e}_R}$:\\[-.8cm]
\bequ
\sigma^{LR}>\sigma^{L0}>\sigma^{00}>\sigma^{LL}>\sigma^{RR}>\sigma^{R0}>\sigma^{RL}
\la{eq_susy1c}
\eequ
\end{itemize} 
If only the electron is polarized then the orderings of the cross sections
for the cases (\re{eq_susy1a}) and (\re{eq_susy1c}) are equal. 
If, however, both beams are simultaneously polarized
the three cases could be distinguished \ci{Gudi_Char_Neut}.

If the selectrons are too heavy to 
be produced in the direct channel one
may still be able to constrain the selectron masses by
studying forward--backward asymmetries, $A_{FB}$,
in the production and decay of neutralinos \ci{Gudi_Char_Neut}. 
Neutralinos are Majorana particles, so that if 
CP is conserved, then neutralino production is exactly
forward--backward symmetric. However, 
due to spin correlations between production and decay,
$A_{FB}\neq 0$ for the decay electron can occur,
 e.g.\, in the reactions $e^+ e^-\to \tilde{\chi}^0_1\tilde{\chi}^0_2$, 
$\tilde{\chi}^0_2\to \tilde{\chi}^0_1 e^+ e^-$ \ci{GMP2}.
Beam polarization enlarges these asymmetries by about a factor 3 if
both beams are simultaneously polarized with (85,60) (Figure~\re{fig_susy3}a). 
In this case the magnitude of $A_{FB}$ can be as large as $13\%$, wheras  
in the case of unpolarized positrons, only 
$4\%$ is possible. Moreover, for a given electron polarization even the sign
of $A_{FB}$ can change when using polarized positrons.
The $A_{FB}$ of the neutralino decay products are sensitive to the selectron 
masses $m_{\tilde{e}_L}$, $m_{\tilde{e}_R}$ as
can be seen in Figure~\re{fig_susy3}b. If direct production of 
selectron pairs is kinematically not accessible, studying $A_{FB}$ allows
constraining the selectron masses. 
Since in the configuration (LR) both the cross sections 
and the asymmetries are larger than those for (L0)
the simultaneous polarization of both beams increases
the accuracy of indirectly determined 
masses of the particles.

Beam polarization is also helpful for constraining the $M_1$ parameter
of the MSSM \ci{Gudi_Char_Neut}.
As can be seen in Figures~\ref{fig_susy4}a,b the cross section,
e.g. $\sigma(e^+ e^-\to \tilde{\chi}^0_1 \tilde{\chi}^0_2) \times
BR(\tilde{\chi}^0_2\to \tilde{\chi}^0_1 e^+ e^-) $, as well as the $A_{FB}$ of 
the decay electron are very sensitive to the $M_1$ parameter.
In regions of the parameter space where it is possible to 
determine $M_1$ by measuring the
neutralino masses \ci{Kneur} this method would provide an independent  
check. Figure~\ref{fig_susy4}a shows that when both beams are polarized
(85,60)
the cross sections increase by  a factor of about 1.6 and the 
statistical error is reduced by a factor of about 0.8 compared to the case 
when 
only the electron beam is polarized. This error reduction would improve 
the accuracy of $A_{FB}$ by about 20\%, Figure~\re{fig_susy4}b.

The MSSM contains 4 neutralinos. One additional
Higgs singlet yields the (M+1)SSM, with 5 neutralinos. 
Superstring--inspired E$_6$--models with even more 
neutral gauge bosons or Higgs singlets 
have a spectrum of 6 or more neutralinos. 
In certain regions of the parameter space, where the
lightest neutralino is singlino--like, the same
mass spectra of the light neutralinos are possible in the MSSM, 
NMSSM and E$_6$.
Beam polarization is sensitive to the different couplings, and can be used to 
distinguish between possible models  \ci{Hesselbach,Franke}.
For example, Figure~\re{fig_susy1}a shows that 
the models can 
be separated by comparing the cross sections.
If only the electron is polarized the difference between the MSSM
cross section of about 6 fb and the (M+1)SSM cross section of 7.5 fb is 
rather small.
For $(+80,-60)$ this difference would be 
larger by more than 
a factor 2. In this case
the cross sections are about 7~fb in the MSSM and up to about 10.5~fb
in the (M+1)SSM, so that it would become easier to distinguish the models. 

In the E$_6$ model the cross sections show 
the same dependence on beam polarization 
\ci{Hesselbach} but they are much smaller
due to the large singlino component in the LSP. One gets less than
1.5~fb in the unpolarized case and less than about 3~fb if only the electron
beam is polarized. However, with $(+80,-60)$
one can reach 5~fb, which could be decisive for measuring the process
(Figure~\re{fig_susy1}b).
This example also shows that polarizing the $e^+$ beams not only increases the
effective polarization but can be used to significantly
enhance the event rates.
\subsubsection{Polarization effects in R--parity violating SUSY}
In R--parity violating SUSY, processes can occur which prefer the 
extraordinary (LL) or (RR) polarization configurations.
An interesting example is $e^+ e^-\to \tilde{\nu} \to e^+ e^-$. 
The main background to this process is Bhabha scattering. 
Polarizing both electrons and positrons can strongly enhance the signal.
A study \ci{Spiesberger} was made for $m_{\tilde{\nu}}=650$~GeV,
$\Gamma_{\tilde{\nu}}=1$~GeV, with an angle cut of 
$45^{0}\le \Theta\le 135^{0}$ and a lepton--number violating coupling 
$\lambda_{131}=0.05$ in the 
R--parity violating Langrangian
${\cal L}_{\not R}\sim \sum_{i,j,k}\lambda_{ijk} L_i L_j E_k$. 
Here $L_{i,j}$ 
denotes the left--handed lepton and squark superfield and $E_k$ the
corresponding right--handed field \ci{Spiesberger}.

\begin{figure}
\begin{center}
\begin{picture}(12,7)
\put(-1.2,0){\includegraphics{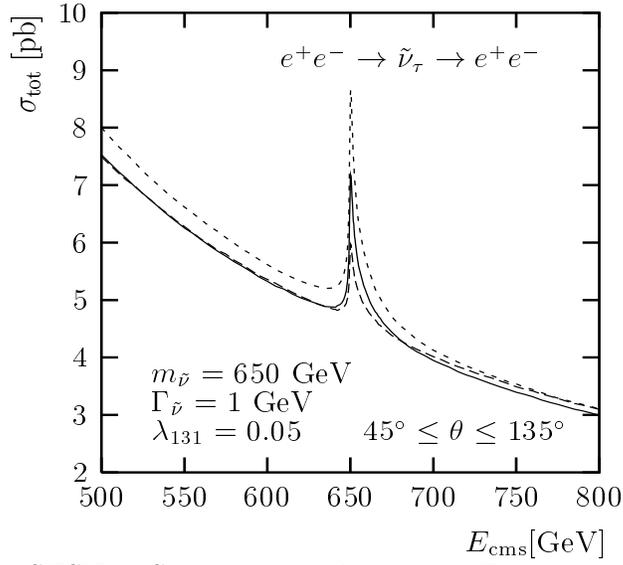}}
\end{picture}\par\vspace{-1.7cm}
\caption{SUSY -- Sneutrino production in R--parity violating model:
Resonance production of
$e^+ e^-\to \tilde{\nu}$ interfering with Bhabha scattering 
for different configurations of 
beam polarization: unpolarized case (solid), 
$P_{e^-}=-80\%$ and $P_{e^+}=+60\%$ (hatched), 
$P_{e^-}=-80\%$ and $P_{e^+}=-60\%$ (dotted) \ci{Spiesberger2}. 
\label{fig_susy5}}
\end{center}
\end{figure}

\bsl
The resonance curve for the process, including the complete 
SM--background is given in Figure~\ref{fig_susy5}. The event rates at the 
peak are given in Table~\ref{tab_susy1}. Electron 
polarization with $(-80,0)$ 
enhances the signal only slightly by about 2\%, whereas the simultaneous
polarization of both beams with $(-80,-60)$ 
produces a further increase by about 20\%. The background changes only 
slightly due to the t--channel (LL) contributions 
from $\gamma$ and $Z$ exchange.
\esl

This configuration of beam polarizations, which strongly suppresses pure SM 
processes, allows one to perform fast diagnostics 
for this R--parity violating process. For example the process
$e^+ e^-\to Z' $ could lead to a similar resonance peak, but with
                                 different polarization dependence. 
Here only 
the `normal' configurations $LR$ and $RL$ play a role and this process will be 
strongly suppressed by $LL$. Therefore such a resonance curve, 
Figure~\ref{fig_susy5}, with different beam polarizations would uniquely
                          identify an 
an R--parity violating SUSY 
process.

\section{Summary}
\label{sec:4}
The clean and fundamental nature of $e^+ e^-$ collisions in a linear 
collider is 
ideally suited for the search for new physics, and the determination of both 
Standard Model and New Physics couplings with high precision.
Polarization effects will play a crucial role in these processes. The fact 
that 
highly polarized electron beams are easily achievable in a linear collider has 
already been demonstrated at the SLAC SLC, and there is every reason to expect
that electron polarizations in excess of $80\%$ will be possible at future 
linear 
colliders. Methods for achieving 40--60\% positron polarization have been 
developed,
and in this paper we have shown that simultaneous polarization of both beams 
can 
significantly expand the accessible physics opportunities. Our main 
conclusions are 
summarized in Table~\ref{tab_sum}, where we list 
the quantitative improvements when simultaneously polarized 
$e^-$ and $e^+$ beams are used
compared to the case where only polarized electrons 
but unpolarized positrons are used.

A recurring theme in this paper is that the simultaneous polarization of both 
electrons and positrons can be used to determine
 quantum numbers of new particles,
increase rates, decrease background, raise the effective polarization,
reduce the error in determining the effective polarization, distinguish 
between 
competing interaction mechanism, and expand the range of measurable 
experimental 
observables. These virtues help to provide us with unique new insights into 
Higgs, 
Electroweak, QCD, SUSY processes and Alternative Theories.

We have shown that polarization is an essential ingredient in the
 determination of 
Higgs couplings. In the electroweak sector, the use of polarized beams should 
make 
possible an order--of--magnitude reduction in the error of the weak mixing 
angle
over the present value. Such a measurement would give us unprecedented 
sensitivity
to radiative corrections caused by processes at otherwise unattainable energy 
scales (GigaZ). 
In QCD polarization influences quark production and improves considerably
the accuracy in determining e.g.\ the top couplings. With highly 
polarized $e^+$ and 
$e^-$ it should be possible to establish polarized photon structure 
functions.
In the search for new gauge bosons, contact interactions and extra dimension
polarization is crucial to enhance rates and suppress background. Finally,
in the yet undiscovered
world of SUSY interactions critical polarization effects abound e.g.\ for the 
discovery and especially
in the determination of the basic coupling parameters. 
Non--standard polarization
configurations can be used to determine  
fundamental quantum numbers of SUSY particles.
For all of these reasons, and perhaps some that we havn't thought of yet, 
polarized 
electron and positron beams should be an integral part of future 
linear colliders.

We would like to thank M. Battaglia, T. Behnke, A. Brandenburg, 
A. De Roeck, K. Desch, M. Diehl, H. Eberl, 
K. Fl\"ottmann, H. Fraas, F. Franke, N. Ghodbane, S. Hesselbach, R. Heuer,
J. Kalinowski, W. Kilian, B. Kniehl,
S. Kraml, J. K\"uhn, W. Majerotto, U. Martyn, W. Menges, K. M\"onig, T. Ohl,
S. Riemann, P. Sch\"uler, R. Settles,
H. Spiesberger,  M. Stratmann, A. Vogt, N. Walker, 
G. Weiglein, G. Wilson, P. Zerwas for interesting discussions.

\begin{table} 
{\small \hspace{-2.9cm}
\parbox{17.8cm}{\caption{
Summary -- polarization effects in all working groups. 
The improvement 
factors indicate the gain if one polarizes simultaneously $e^-$ and $e^+$ 
beams and refer to the polarizations given in the text, 
e.g. (80,60), compared to the case when only $e^-$ is polarized, e.g. (80,0),
respectively.
\la{tab_sum}} }\vspace{.2cm}
\hspace*{-3cm}\begin{tabular}{|l|l|ll|}
\hline
Process & Background & $P(e^+)$  Improvement Factors&  \\ \hline\hline
Higgs &&&\\ \hline
$e^+ e^-\to H \bar{\nu} \nu$& $WW$, $ZZ$, $Z\bar{\nu} \nu$ & Enhancing of
$\frac{S}{B}$, $\frac{S}{\sqrt{B}}$ & factor 1.2--1.3 \\
$e^+ e^-\to HZ$ & & better separation: $HZ\leftrightarrow H\bar{\nu}\nu$ & 
factor 4 with RL\\
 & $W^{\pm} \ell^{\mp} \stackrel{(-)}{\nu}$ & suppression of single $W$ & 
important\\
$e^+e^-\to HZ\to H \bar{f} f$ &  & general HZV & error reduction by $30\%$ \\ 
\hline\hline
Electroweak &&& \\ \hline
high $\sqrt{s}$: &&&\\
$e^+ e^-\to W^+ W^-$ & $\gamma \gamma \to W^+ W^-$ & 
Enhancing of $\frac{S}{B}$, $\frac{S}{\sqrt{B}}$ & up to a factor 2 \\ 
 & $W^{\pm} Z \nu$, $W^+ W^- Z$ 
& $\Delta \kappa_{\gamma}$, $\Delta \lambda_{\gamma}$, $\Delta \kappa_{Z}$, 
$\Delta \lambda_{Z}$,   & error reduction by a factor 1.8 \\
Giga Z: &&&\\
$e^+ e^-\to Z$ & & improve $\delta(P)$, $A_{LR}$ & factor 10  \\ 
\hline\hline
QCD &&& \\ \hline
$e^+ e^-\to q \bar{q}$ & $e^+ e^-\to W^+ W^-$ & 
Enhancing of $\frac{S}{B}$, $\frac{S}{\sqrt{B}}$ & factor 2--3 with RL \\
$e^+ e^-\to \gamma \gamma\to jets$ &
 & pol. structure functions& high $P(e^+)$ essential\\
$e^+ e^-\to t \bar{q}$ & $e^+ e^-\to W q \bar{q}'$ 
 & Enhancing of $\frac{S}{B}$, $\frac{S}{\sqrt{B}}$ & factor 2\\ 
&& $Vtq$ couplings (FCNC) &
limits reduction by 40\% 
\\ \hline
\hline
SUSY  &&& \\ \hline
$e^+ e^-\to \tilde{\ell} \tilde{\ell}$, $\tilde{\chi}^+\tilde{\chi}^-$,
$\tilde{\chi}^0 \tilde{\chi}^0$ &
$W^+ W^-$, $ZZ$  &better separation   
$\tilde{\ell}_L \leftrightarrow \tilde{\ell}_R$ & factor 1.6 \\ 
 &&Test of quantum numbers $L$, $R$& uniquely with $LL$, $RR$\\
 && better indirect $m_{\tilde{\nu}}$ & \\
 && separation between models & important\\
 && Enhancing of $\frac{S}{B}$, $\frac{S}{\sqrt{B}}$ & factor 2--3\\
 && error reduction in $M$, $\mu$, $\tan\beta$ & \\ 
$e^+ e^-\to \tilde{\chi}^{\pm} \tilde{e} \nu$ & 
& enhancement with LL, RR & probably essential\\ \hline \hline
Alternative Theories &&& \\ \hline
$e^+e^-\to \gamma,Z,Z',W'W'$ & $W^+ W^-$, $W^{\pm} e^{\mp} \nu$,
& Enhancing of $\frac{S}{B}$, $\frac{S}{\sqrt{B}}$
 & model dependent \\
 & $WWZ$, $f \bar{f}$, $ZH$ && \\
&& discovery reach of $W'$, $Z'$ & enlarged by about $10\%-20\%$  \\
$\not R$: $e^+ e^-\to \tilde{\nu}_{\tau}\to e^+ e^-$ & $e^+ e^-$ & Enhancing 
$S/B$, $S/\sqrt{B}$ &
 factor 10 with $LL$ \\
 && Test of quantum number & \\
CI: $e^+e^-\to \mu^+ \mu^-$, $q \bar{q}$ & & sensitivity & enlarged by 
$40\%$, O($10^2$ TeV) \\
ED: $e^+ e^-\to \gamma G$ & $e^+ e^-\to \nu \bar{\nu} \gamma $ &
Enhancing of $\frac{S}{B}$, $\frac{S}{\sqrt{B}}$ & up to a factor 5\\ 
&& discovery reach  & enlarged by $20\%$, O(TeV) \\
\hline
\end{tabular}\par
}
\end{table}
 
\begin{table}\hspace{-2cm}
\parbox{16.5cm}{\caption{Higgs production in Standard Model: 
Scaling factors, i.e. ratios of polarized and unpolarized
cross section $\sigma^{pol}/\sigma^{unpol}$,
are given in Higgs production and background processes 
for different polarization configurations with
$|P_{e^-}|=80\%$, $|P_{e^+}|=60\%$ \ci{Desch_obernai}. 
\label{tab_higgs1}}}\vspace{.2cm}
\hspace*{-2cm}\begin{tabular}{|l||c|c||c|c|}
\hline
Configuration 
& \multicolumn{2}{|c||}{Higgs Production} &
\multicolumn{2}{||c|}{Background} \\
$(sgn(P_{e^-}) sgn(P_{e^+}))$
&$e^+e^-\to H \nu \bar{\nu}$ & $e^+ e^-\to HZ$ & 
$e^+e^-\to WW$, $e^+e^-\to Z \nu \bar{\nu}$ & $e^+e^-\to ZZ$ \\ \hline
$(+0)$ & 0.20 & 0.87 & 0.20 & 0.76 \\
$(-0)$ & 1.80 & 1.13 & 1.80 & 1.25 \\ \hline
$(+-)$ & 0.08 & 1.26 & 0.10 & 1.05 \\
$(-+)$ & 2.88 & 1.70 & 2.85 & 1.91 \\ \hline
\end{tabular}
\end{table}

\begin{table}\hspace{-2.5cm}
\parbox{17.5cm}{\caption{Higgs production -- Determination of
general Higgs couplings: Optimal errors on general $ZZ\Phi$ and 
$Z\gamma \Phi$ couplings for different efficiencies $\epsilon_{\tau}$, 
$\epsilon_b$
and beam polarizations
\ci{Kniehl_desy00}. \label{tab_higgs2} }}\vspace{.2cm}
\hspace*{-2.5cm}\begin{tabular}{|l||c|c|c|c|}
\hline
 & \multicolumn{3}{|c|}{$\epsilon_{\tau}=0=\epsilon_b$}
& $\epsilon_{\tau}=50\%$, $\epsilon_b=60\%$ \\ \hline
 & $P_{e^-}=0=P_{e^+}$ & $P_{e^-}=80\%$, $P_{e^+}=0$ &
$P_{e^-}=80\%$, $P_{e^+}=60\%$ & $P_{e^-}=80\%$, $P_{e^+}=60\%$ 
\\ \hline
Re($b_Z$) & 0.00055 & 0.00028 & 0.00023 & 0.00022 \\
Re($c_Z$) & 0.00065 & 0.00014 & 0.00011 & 0.00011 \\ \hline
Re($b_{\gamma}$) & 0.01232 & 0.00052 & 0.00036 & 0.00034 \\
Re($c_{\gamma}$) & 0.00542 & 0.00011 & 0.00008 & 0.00007 \\ \hline
Re($\tilde{b}_Z$) & 0.00104 & 0.00095 & 0.00078 & 0.00052 \\
Re($\tilde{b}_{\gamma}$) & 0.00618 & 0.00145 & 0.00101 & 0.00063 \\ \hline
Im($b_Z-c_Z$) & 0.01055 & 0.00070 & 0.00049 & 0.00046 \\
Im($b_{\gamma}-c_{\gamma}$) & 0.00206 & 0.00070 & 0.00057 & 0.00054\\
Im($\tilde{b}_Z$) & 0.00521 & 0.00032 & 0.00022 & 0.00022 \\
Im($\tilde{b}_{\gamma}$) & 0.00101 & 0.00032 & 0.00026 & 0.00026 \\ \hline
\end{tabular}
\end{table}

\begin{table}\hspace{-2cm}
\begin{center}\hspace*{-2cm}
\parbox{16.8cm}{\caption{Electroweak Theory -- Determination of couplings:
Expected sensitivity ($\times 10^{-4}$) for different triple gauge couplings 
$\kappa_V$ and anomalous triple gauge couplings $\lambda_V$
 in $e^+ e^-\to W^+ W^-$
     at a center-of-mass energy of $\sqrt{s}=500$~GeV and
      $\sqrt{s}=800$~GeV. In the case of 
      polarized beams the luminosity is split up equally on both
      combinations; $\delta g_1^{Z1}$, 
$\Delta \kappa_{\gamma}^1$, $\lambda_{\gamma}^1$ 
were evaluated by using $SU(2)\times U(1)$ relations \ci{Menges}.
\label{tab_ew1}} }
\renewcommand{\footnoterule}{}
\hspace*{-2cm}
\begin{tabular}{|c||rrr||rrrrr||rrrr|}
\hline
& $\Cdgz\footnotemark[1]$ & $\Cdkg\footnotemark[1]$ &
$\Clg\footnotemark[1]$ & 
 $\Cdgz$ & $\Cdkg$ & $\Clg$  & $\Cdkz$ & $\Clz$ & $\Cgz{4}$ & $\Cgz{5}$ & 
$\Ckzt$ & $\Clzt$ \\
\hline\hline
\multicolumn{13}{|c|}{unpolarized beams}\\ \hline
$\sqrt{s}=500$~GeV & 7.3 & 5.7 & 6.1 & 38.1 & 4.8 & 12.1 & 8.7 & 11.5 &
85.8 & 27.7 & 64.9 & 11.4 \\
$\sqrt{s}=800$~GeV & 4.5 & 3.1 & 2.8 & 39.0 & 2.6 &  5.2 & 4.9 &  5.1 &
41.8 & 28.5 & 29.6 &  4.9 \\
\hline
\multicolumn{13}{|c|}{only electron beam polarized, 
$|P_{e^-}|=80\%$}\\ \hline
$\sqrt{s}=500$~GeV & 4.3 & 4.2 & 5.1 & 24.8 & 4.1 &  8.2 & 5.0 &  8.9 &
79.9 & 22.8 & 50.6 & 10.3 \\
$\sqrt{s}=800$~GeV & 2.7 & 2.3 & 3.1 & 21.9 & 2.2 &  5.0 & 2.9 &  4.7 &
31.8 & 24.3 & 24.1 &  4.4 \\
\hline
\multicolumn{13}{|c|}{both beams polarized, $|P_{e^-}|=80\%$ and 
$|P_{e^+}|=60\%$}\\ \hline
$\sqrt{s}=500$~GeV & 2.8 & 3.1 & 4.3 & 15.5 & 3.3 &  5.9 & 3.2 &  6.7 &
45.9 & 16.5 & 39.0 &  7.5 \\
$\sqrt{s}=800$~GeV & 1.8 & 1.9 & 2.6 & 12.6 & 1.9 &  3.3 & 1.9 &  3.0 &
18.3 & 14.4 & 14.3 &  3.0 \\
\hline
\end{tabular}
\end{center}
\end{table}

\begin{table}
\begin{center}
\parbox{9.8cm}{\caption{QCD -- Quark production: 
Scaling factors, i.e. ratios of polarized and unpolarized cross 
sections $\sigma^{pol}/\sigma^{unpol}$, 
of light quark production $e^+ e^-\to q \bar{q}$ at 
$\sqrt{s}=500$~GeV \ci{Brandenburg}. \label{tab_qcd1}} }
\begin{tabular}{|l||c|c||c|c|}
\hline
& \multicolumn{4}{|c|}{Configuration of $(sign(P_{e^-}) sign(P_{e^+}))$}\\ 
\hline
scaling factors& \quad$L0$\quad & \quad$LR$\quad & \quad$R0$\quad & \quad$RL$\quad \\ \hline
 $e^+e^-\to q \bar{q}$ & 1.37 & 2.12 & 0.63 & 0.87 \\ \hline
\end{tabular}
\end{center}
\end{table}

\begin{table}
\begin{center}
\hspace*{-.8cm}\parbox{14cm}{\caption{QCD -- Determination of top couplings:
Limits 95\% C.L. of top flavour changing neutral couplings 
from top branching fractions at 
$\sqrt{s}=500$~GeV with ${\cal L}=300$~fb$^{-1}$ and at $\sqrt{s}=800$~GeV
with ${\cal L}=500$~fb$^{-1}$ \ci{Aguilar}. \label{tab_qcd2} }}\vspace{.2cm}
\hspace*{-1cm}
\begin{tabular}{|l||c|c|c|}
\hline 
 & unpolarized beams & $|P_{e^-}|=80\%$ & $|P_{e^-}|=80\%$, $|P_{e^+}|=45\%$ \\
 \hline\hline
 & \multicolumn{3}{|c|}{$\sqrt{s}=500$~GeV}\\ \hline
 $BR(t\to Zq)(\gamma_{\mu})$ & $4.4\times 10^{-4}$ & $3.1\times 10^{-4}$ &
$1.9\times 10^{-4}$ \\
 $BR(t\to Zq)(\sigma_{\mu\nu})$ & $3.5\times 10^{-5}$ & $2.4\times 10^{-5}$ &
$1.5\times 10^{-5}$ \\
$BR(t\to \gamma q)$ & $2.2\times 10^{-5}$ & $1.3\times 10^{-5}$ &
$8.2\times 10^{-6}$ \\ \hline\hline
 & \multicolumn{3}{|c|}{$\sqrt{s}=800$~GeV} \\ \hline
 $BR(t\to Zq)(\gamma_{\mu})$ & $4.4\times 10^{-4}$ & $2.9\times 10^{-4}$ &
$2.4\times 10^{-4}$ \\
 $BR(t\to Zq)(\sigma_{\mu\nu})$ & $1.3\times 10^{-5}$ & $8.6\times 10^{-6}$ &
$6.2\times 10^{-6}$ \\
$BR(t\to \gamma q)$ & $7.8\times 10^{-6}$ & $4.5\times 10^{-6}$ &
$3.7\times 10^{-6}$\\ \hline
\end{tabular}
\end{center}
\end{table}

\begin{table}
\hspace*{-1.9cm}\parbox{17.3cm}{\caption{Alternative Theories -- Reach for additional
gauge bosons $Z'$ and $W'$ in different models:
Lower bounds of $M_{Z'}$ in an E$_6$-- and a $LR$--model for
${\cal L}=500$~fb$^{-1}$ \ci{Riemann}.
Lower bounds of $M_{W'}$ in a model with SM--coupling $W'$, in a LR--model 
and in a Kaluza--Klein model \ci{Godfrey}. \la{tab_alt1} }}\vspace{.2cm}
\hspace*{-2cm}
\begin{tabular}{|l||c|c||c|c|c|}
\hline
 & \multicolumn{2}{|c||}{${\cal L}=500$~fb$^{-1}$} &
\multicolumn{3}{|c|}{${\cal L}=1$~ab$^{-1}$} \\
 & $M_{Z'}$ in E$_6$ & $M_{Z'}$ in LR & $M_{W'}$ in SSM & $M_{W'}$ in LR
& $M_{W'}$ in KK \\ \hline
unpolarized case & -- & -- & 1.7 TeV & 0.9 TeV & 1.8 TeV \\
$|P_{e^-}|=80 \%$ & 4.4 TeV & 4.0 TeV & 2.2 TeV & 1.2 TeV &  2.3 TeV \\
$|P_{e^-}|=80 \%$, $|P_{e^+}|=60 \%$ & 4.9 TeV & 6.0 TeV & 
\multicolumn{3}{|c|}{rough approximation: further gain about 20\%}\\\hline   
\end{tabular}
\end{table}

\begin{table}
\hspace*{-.9cm}\parbox{14.2cm}{\caption{\label{tab_alt2}
Alternative Theories -- Search for extra dimensions in graviton emission:
Scale $M_{D}$/TeV in dependence of the numbers $\delta$ of extra dimensions
and different configurations of beam polarizations 
for $\sqrt{s}=800 \,\mbox{GeV}$ and ${\cal L}_{int}=1000 \,\mbox{fb}^{-1}$ 
at a confidence level of
95 \% ($\Delta\chi^2 =3.84$) for the normalisation uncertainties
$\frac{\Delta f_N}{f_N} = 1\,\%$ 
and for $\frac{\Delta f_N}{f_N} = 0.1\,\%$ \ci{Vest}.} }\vspace{.2cm}
\hspace*{-1cm}
    \begin{tabular}[t]{|l||l||r|r|}
      \hline 
        & & $M_D$/TeV & Scale $M_D$/TeV \\
       Number of & & for C.L.= 95 \% & for C.L.= 95 \% 
       \\ Extra Dimensions & Polarisation & and 
$\frac{\Delta f_N}{f_N} = 1\,\%
$ & and $\frac{\Delta f_N}{f_N} = 0.1\,\%$ \\
      \hline 
      $\delta=2$ & $P_{e_R^-}=0$, $P_{e_L^+}=0$ & 6.0 & 6.5 \\
      & $P_{e_R^-}=0.8$, $P_{e_L^+}=0$ & 7.4 & 8.7 \\
      & $P_{e_R^-}=0.8$, $P_{e_L^+}=-0.45$ & 8.7 & 10.0 \\
      & $P_{e_R^-}=0.8$, $P_{e_L^+}=-0.6$ & 9.4 & 10.8 \\
      \hline 
      $\delta=3$ & $P_{e_R^-}=0$, $P_{e_L^+}=0$ & 4.4 & 4.8 \\
      & $P_{e_R^-}=0.8$, $P_{e_L^+}=0$ & 5.2 & 6.0 \\
      & $P_{e_R^-}=0.8$, $P_{e_L^+}=-0.45$ & 6.0 & 6.8 \\
      & $P_{e_R^-}=0.8$, $P_{e_L^+}=-0.6$ & 6.4 & 7.3 \\
      \hline 
      $\delta=4$ & $P_{e_R^-}=0$, $P_{e_L^+}=0$ & 3.6 & 3.7 \\
      & $P_{e_R^-}=0.8$, $P_{e_L^+}=0$ & 4.2 & 4.6 \\
      & $P_{e_R^-}=0.8$, $P_{e_L^+}=-0.45$ & 4.6 & 5.0 \\
      & $P_{e_R^-}=0.8$, $P_{e_L^+}=-0.6$ & 4.8 & 5.3 \\
      \hline 
      $\delta=5$ & $P_{e_R^-}=0$, $P_{e_L^+}=0$ & 2.9 & 3.1 \\
      & $P_{e_R^-}=0.8$, $P_{e_L^+}=0$ & 3.3 & 3.6 \\
      & $P_{e_R^-}=0.8$, $P_{e_L^+}=-0.45$ & 3.6 & 3.9 \\
      & $P_{e_R^-}=0.8$, $P_{e_L^+}=-0.6$ & 3.8 & 4.2 \\
      \hline 
      $\delta=6$ & $P_{e_R^-}=0$, $P_{e_L^+}=0$ & 2.5 & 2.6 \\
      & $P_{e_R^-}=0.8$, $P_{e_L^+}=0$ & 2.8 & 3.1 \\
      & $P_{e_R^-}=0.8$, $P_{e_L^+}=-0.45$ & 3.1 & 3.3 \\
      & $P_{e_R^-}=0.8$, $P_{e_L^+}=-0.6$ & 3.2 & 3.4 \\
      \hline 
      $\delta=7$ & $P_{e_R^-}=0$, $P_{e_L^+}=0$ & 1.9 & 2.0 \\
      & $P_{e_R^-}=0.8$, $P_{e_L^+}=0$ & 2.2 & 2.3 \\
      & $P_{e_R^-}=0.8$, $P_{e_L^+}=-0.45$ & 2.3 & 2.4 \\
      & $P_{e_R^-}=0.8$, $P_{e_L^+}=-0.6$ & 2.4 & 2.6 \\
     \hline
    \end{tabular}
\end{table}

\begin{table}
\begin{center}
\hspace*{.1cm}\parbox{12.8cm}{\caption{SUSY -- Sneutrino production in 
R--parity violating SUSY:
Cross sections of $e^+ e^-\to \tilde{\nu}\to e^+ e^-$ for 
  unpolarized beams, $P_{e^-}=-80\%$ and unpolarized positrons and
$P_{e^-}=-80\%$, $P_{e^+}=-60\%$. The study was made for 
$m_{\tilde{\nu}}=650$~GeV, $\Gamma_{\tilde{\nu}}=1$~GeV, an angle cut
of $45^0\le \theta \le 135^0$ and the R--parity violating coupling 
$\lambda_{131}=0.05$ \ci{Spiesberger2}.  
\label{tab_susy1}} }\vspace{.2cm}
\begin{tabular}{|l||c|c|}
\hline
 & $\sigma(e^+e^-\to e^+ e^-)$ with &  Bhabha--background\\ 
& $\sigma(e^+ e^-\to \tilde{\nu}\to e^+ e^-)$&  \\ \hline
unpolarized & 7.17 pb & 4.50 pb \\
$P_{e^-}=-80\%$ & 7.32 pb & 4.63 pb\\  
$P_{e^-}=-80\%$, $P_{e^+}=-60\%$ & 8.66 pb & 4.69 pb\\
$P_{e^-}=-80\%$, $P_{e^+}=+60\%$ & 5.97 pb  & 4.58 pb\\
\hline 
\end{tabular}
\end{center}
\end{table}

\begin{figure}[t]
\hspace{.9cm}
\begin{center}
\begin{picture}(9,5)
\put(-.2,0){\includegraphics{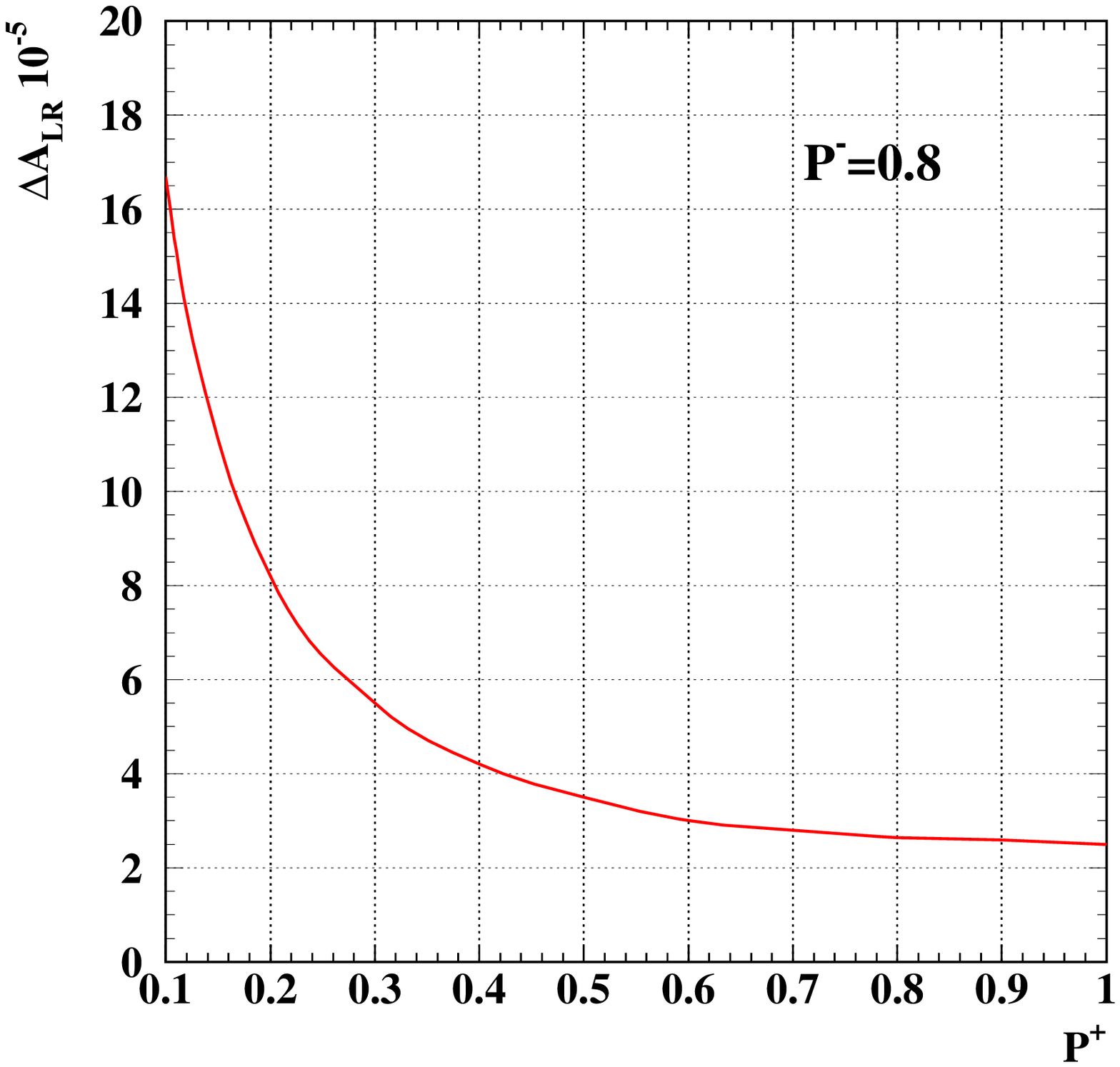}}
\end{picture}
\caption{ Test of Electroweak Theory: 
The statistical error on the left--right asymmetry 
$A_{LR}$ of $e^+ e^-\to Z\to \ell \bar{\ell}$ at GigaZ 
as a function of the positron 
polarization $P(e^+)$ 
for fixed electron polarization $P_{e^-}=\pm 80\%$
\ci{Moenig}.\la{fig_ew1}}
\vspace{2cm}
\end{center}
\hspace{-.9cm}
\begin{center}
\begin{picture}(9,5)
\put(-.2,0){\includegraphics{fig2.eps}}
\end{picture}
\caption{Test of Electroweak Theory: 
A high-precision measurement at GigaZ of the left--right asymmetry
$A_{LR}$ and 
consequently of $\sin^2\Theta_{eff}^l$ 
allows to test the electroweak theory at an unprecedented
level. The allowed parameter space of the SM and the MSSM
in the $\sin^2\Theta_{eff}^l$--$M_W$ plane is shown together with the 
experimental
accuracy reachable at GigaZ. For comparison, the present experimental
accuracy (LEP/SLD/Tevatron) and the prospective accuracy at the LHC
and a LC without GigaZ option (LHC/LC) are also shown \ci{Weiglein2}.  \la{fig_ew2}}
\end{center}
\end{figure}

\begin{figure}[t]
\vspace{-1.8cm}
\begin{center}
\begin{minipage}{7cm}
\begin{picture}(7,5)
\put(-.2,0){\includegraphics{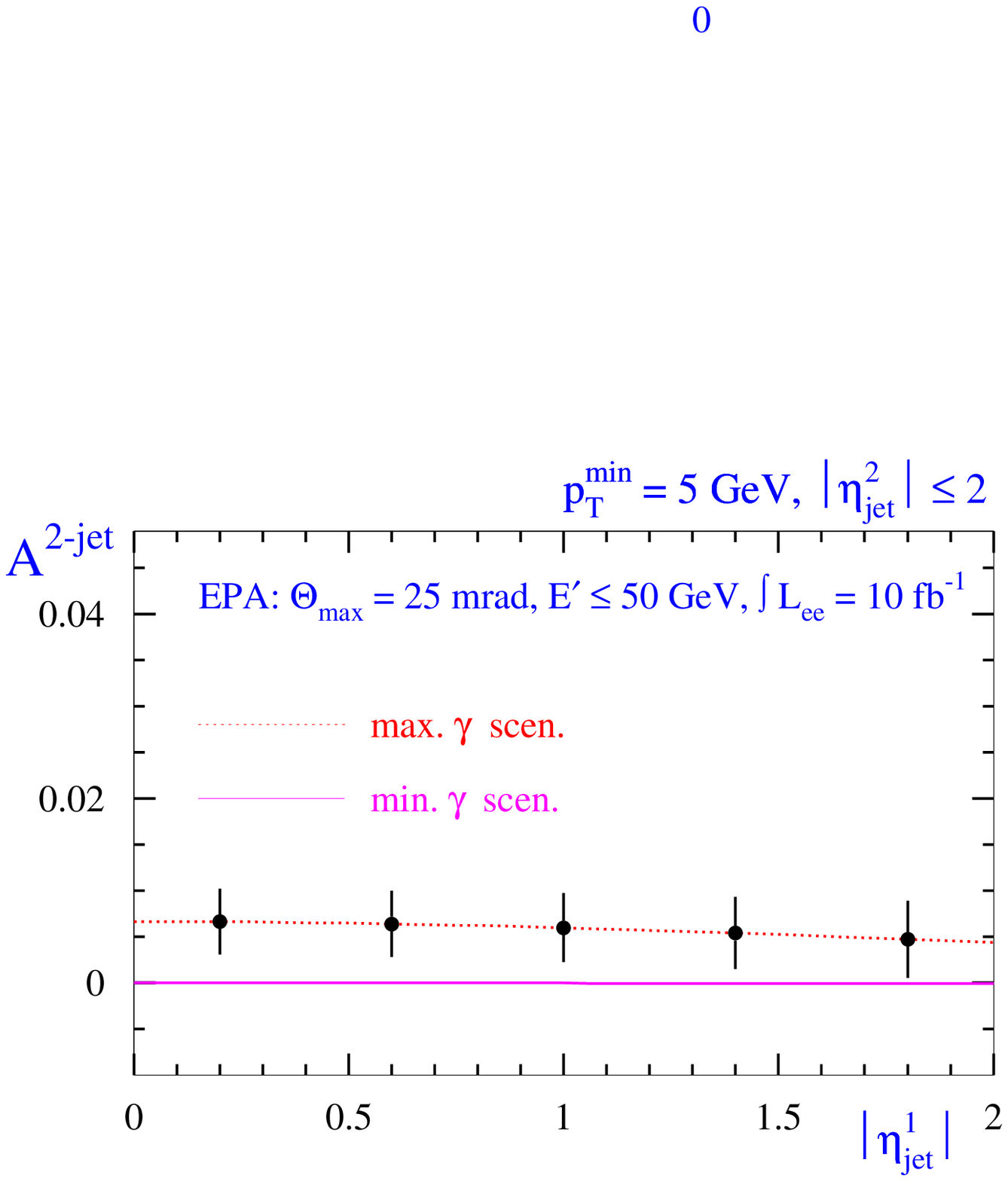}}
\end{picture}\par\vspace{.9cm}
\end{minipage}
\caption{QCD -- Establishing of polarized photon structure functions:
Di--jet asymmetry of $e^+ e^-\to e^+ e^- +$Di--jets
for events with $p_T^{jet}=5$~GeV
and $|\eta_{jet}|<2$ for $\gamma\gamma$ collisions at an $e^+ e^-$ collider.
Since only ${\cal L}_{ee}=10$~fb$^{-1}$ the statistical 
errors of the small di--jet 
asymmetries are quite large, $\sim 0.5 \%$. If, however, 
${\cal L}_{ee}=100$~fb$^{-1}$ is assumed, which would be reachable at TESLA,
the errors are reduced by about a factor 3. 
Optimization of cuts will further reduce the errors \ci{Stratmann}.
\la{fig_qcd1}}
\end{center}
\vspace{.1cm}
\begin{center}
\begin{picture}(12,8)
\put(-.2,0){\includegraphics{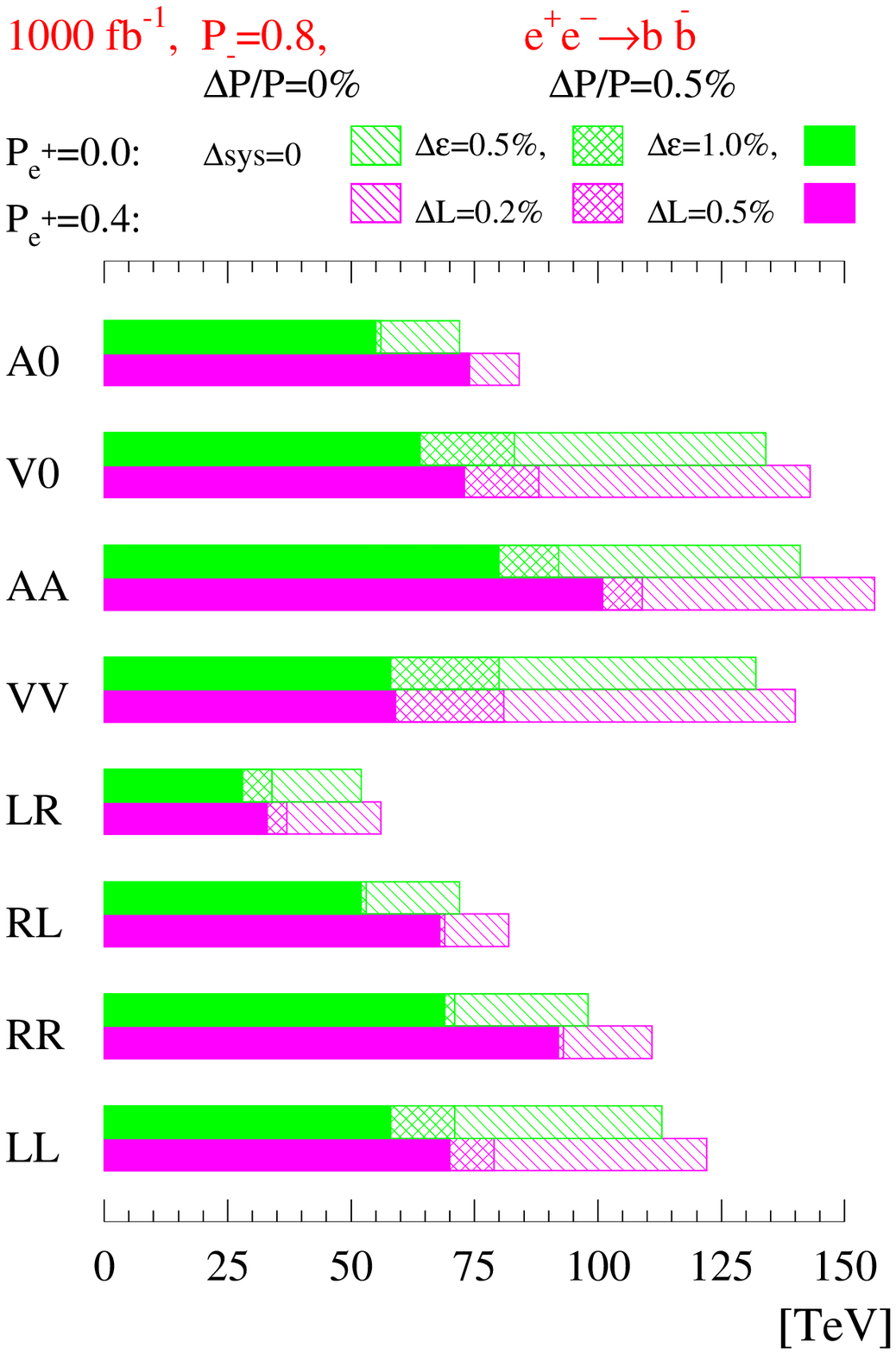}}
\put(6,-1){\small $\Lambda$}
\end{picture}\par\vspace{.7cm}
\caption{Alternative Theories -- Reach for different contact interactions:
Expected sensitivities (95\%) to contact interaction scales $\Lambda$
in $e^+ e^-\to b \bar{b}$ at $\sqrt{s}=500$~GeV with ${\cal L}=1000$~fb$^{-1}$
for $P_{e^-}=80\%$ and unpolarized positrons and both beams polarized with
$|P_{e^-}|=80\%$, $|P_{e^+}|=40\%$. \ci{Riemann}.
 \la{fig_alt1}}
\end{center}
\end{figure}

\begin{figure}
\vspace{-1cm}
\begin{center}
\begin{picture}(12,8)
\put(1,0){\includegraphics{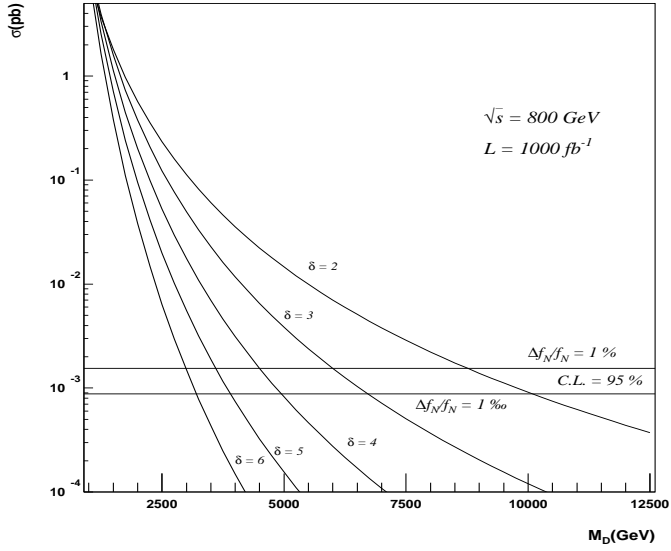}}
\end{picture}\par\vspace{-1.5cm}
\caption{Alternative Theories -- Search for extra dimensions in graviton emission:
Total cross sections for $e^+ e^-\to \gamma G$ with $P_{e^-}=80\%$
and $P_{e^+}=60\%$ at
$\sqrt{s}=800$~GeV, ${\cal L}=1000$~fb$^{-1}$ 
as a function of the scale $M_D$ [TeV] 
for different numbers $\delta$ of extra dimensions. The two horizontal 
lines indicate two different normalization
uncertainties $\frac{\Delta f_N}{f_N}=1\%$ 
and $\frac{\Delta f_N}{f_N}=0.1\%$  \ci{Vest}. \la{fig_alt2}}
\vspace{2cm}\hspace*{1.5cm}
\end{center}
\end{figure}

\begin{figure}
\hspace{2cm}
\begin{picture}(15,8)
\put(-.2,0){\includegraphics{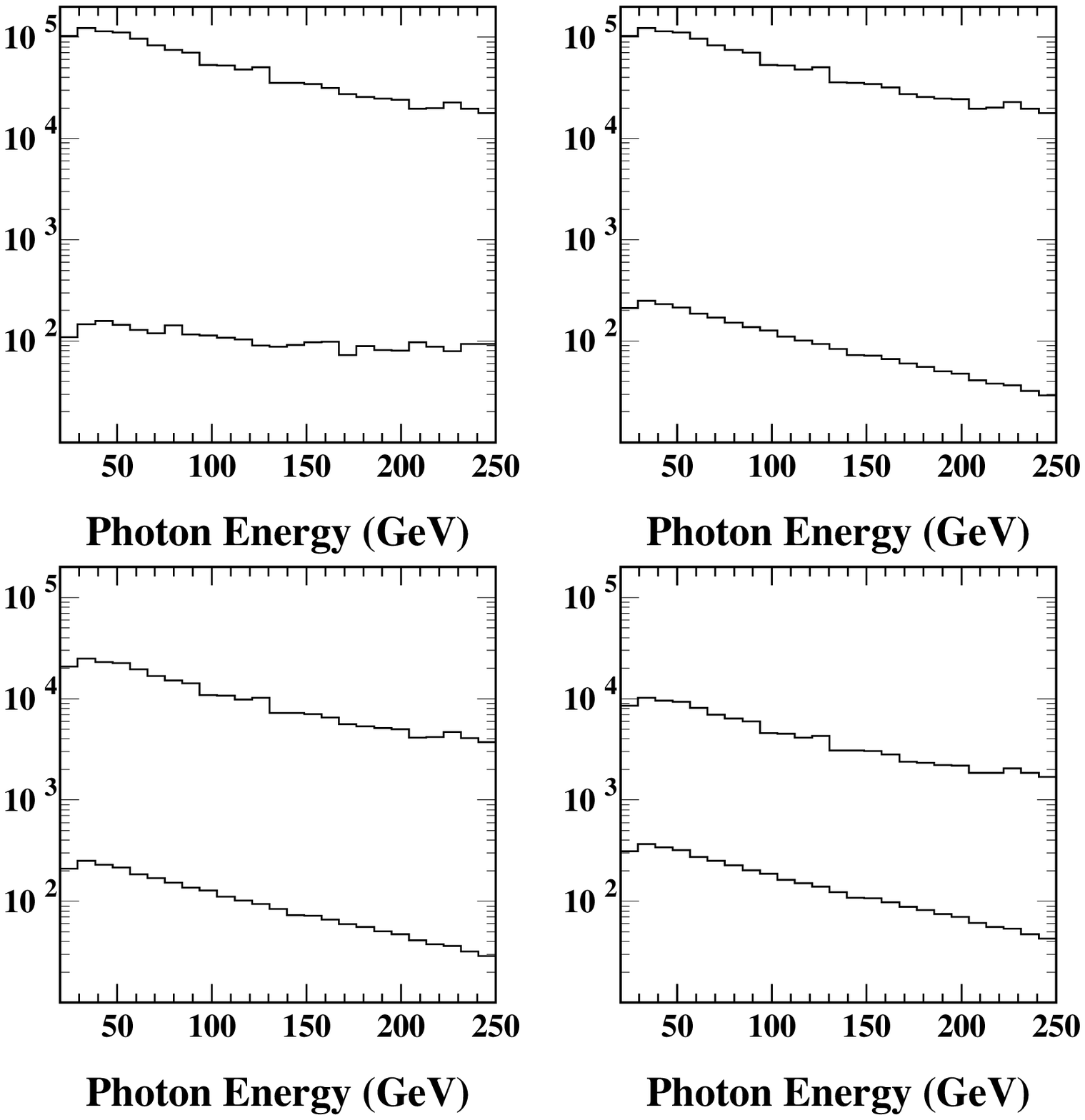}}
\put(.3,8.5){\small $\sigma_L$}
\put(.3,6.2){\small $\sigma_R$}
\put(.9,7.4){$e^+ e^-\to \nu \bar{\nu} \gamma$}
\put(.2,3.4){\small $e^+ e^-\to \nu \bar{\nu} \gamma$}
\put(.2,1.7){\small $e^+ e^-\to \gamma G$}
\put(2,4.2){\small $P_{e^-}=80\%$}
\put(5.4,3.1){\small $e^+ e^-\to \nu \bar{\nu} \gamma$}
\put(5.4,1.8){\small $e^+ e^-\to \gamma G$}
\put(5.4,4.2){\small $P_{e^-}=80\%$, $P_{e^+}=60\%$}
\put(5.5,8.3){\small $e^+ e^-\to \nu \bar{\nu} \gamma$}
\put(5.5,6.15){\small $e^+ e^-\to \gamma G$}
\put(6.8,8.8){\small $P_{e^-}=0=P_{e^+}$}
\end{picture}\par\vspace{-.7cm}
\hspace*{-1cm}\parbox{15cm}{\caption{Alternative Theories -- Search for 
extra dimensions in graviton emission:
Background $e^+ e^-\to \nu \bar{\nu} \gamma$ to direct search 
$e^+ e^-\to \gamma G$ at $\sqrt{s}=800$~GeV, ${\cal L}=1000$~fb$^{-1}$
for $\delta=2$, $M_D=7.5$~TeV 
as function of the photon energy for different 
configurations of beam polarizations. With both beams polarized, 
$P_{e^-}=80\%$ and $P_{e^+}=-60\%$ the significance $S/\sqrt{B}$ is 
improved by a factor 5, with 
electron polarization but unpolarized positron one gains only
the factor 2.2 \ci{Wilson}.  
\la{fig_alt3}}}\par\vspace{1.4cm}
%
\hspace{-.7cm}
\begin{minipage}{7cm}
\begin{picture}(7,5)
\put(-.2,0){\includegraphics{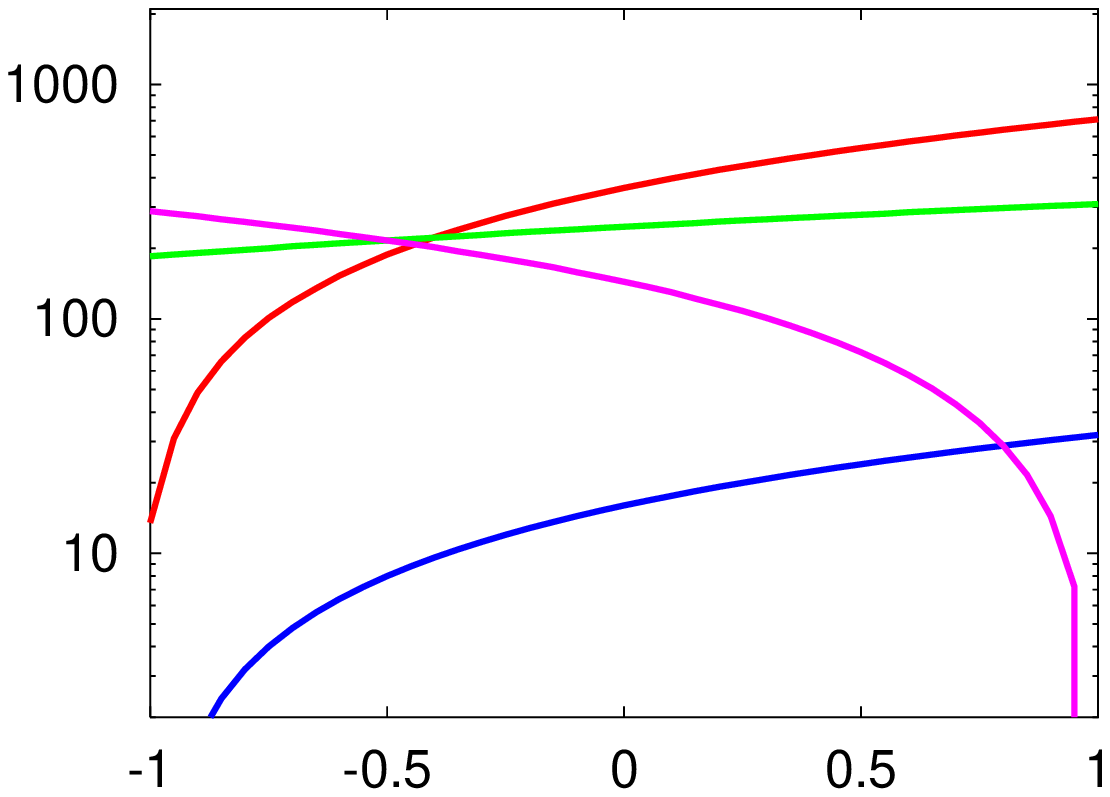}}
\put(-1.4,5.5){\Large $\sigma$/[fb]}
\put(6,-.8){\Large $P_{e^+}$}
\put(1.8,5.5){\Large\color{Black} $P_{e^-}=-80\%$}
\put(3.7,.5){\large $\sqrt{s}=500$~GeV}
\put(4.5,4.6){\Large\color{Red} $\tilde{e}^+_L\tilde{e}^-_L$}
\put(5.2,3.3){\Large\color{Olive} $\tilde{e}^+_R \tilde{e}^-_R$}
\put(.3,4.1){\Large\color{Lila} $\tilde{e}^+_R \tilde{e}^-_L$}
\put(.8,1.6){\Large\color{Blue} $\tilde{e}^+_L \tilde{e}^-_R$}
\end{picture}\par\vspace{+.7cm}
\end{minipage}\hfill\hspace{.2cm}
\begin{minipage}{7cm}
\begin{picture}(7,5) 
\put(-.2,0){\includegraphics{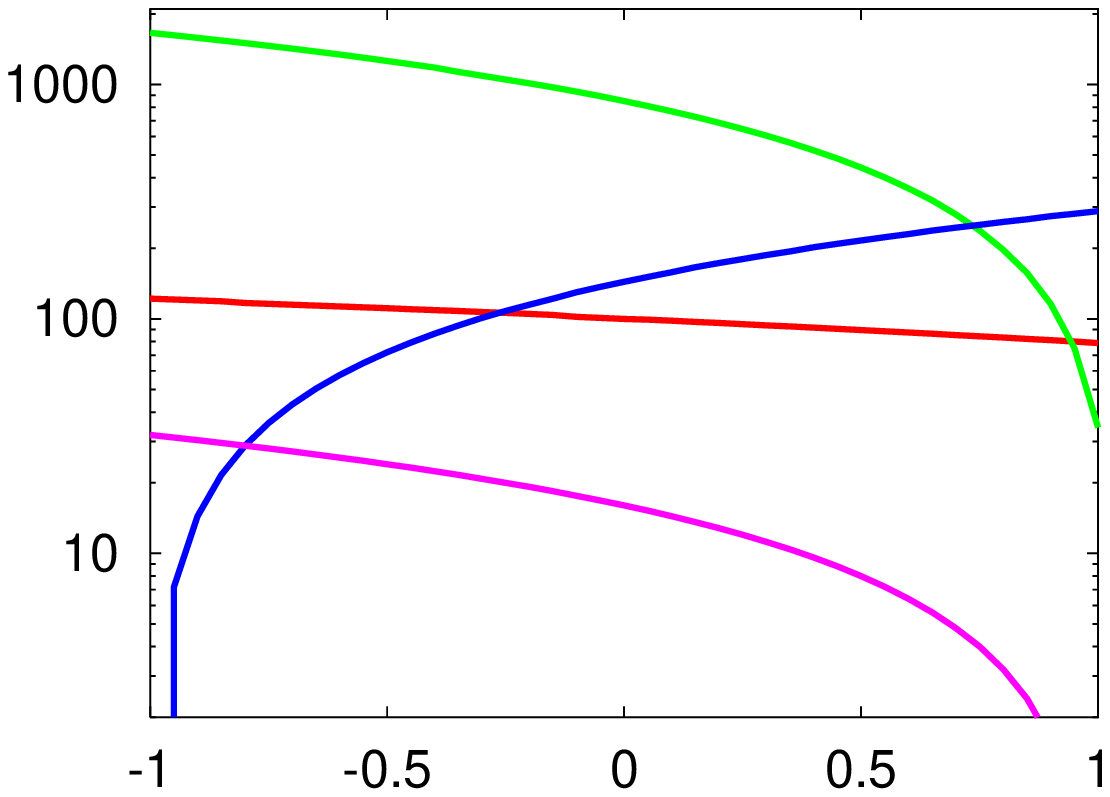}}
\put(-.5,5.5){\Large $\sigma$/[fb]}
\put(6.8,-.8){\Large $P_{e^+}$}
\put(2.5,5.5){\Large\color{Black} $P_{e^-}=+80\%$}
\put(1.1,.5){\large $\sqrt{s}=500$~GeV}
\put(1.2,3.4){\Large\color{Red} $\tilde{e}^+_L\tilde{e}^-_L$}
\put(5.1,4.6){\Large\color{Olive} $\tilde{e}^+_R \tilde{e}^-_R$}
\put(4.8,1.8){\Large\color{Lila} $\tilde{e}^+_R \tilde{e}^-_L$}
\put(3.7,3.7){\Large\color{Blue} $\tilde{e}^+_L \tilde{e}^-_R$}
\end{picture}\par\vspace{.7cm}
\end{minipage}
\hspace*{-1cm}\parbox{15cm}{\caption{SUSY -- Slepton production: Cross 
sections for 
$e^+ e^-\to \tilde{e}_{L,R}\tilde{e}_{L,R}$ [fb] in the reference scenario
\ci{Blair} are shown
  for fixed electron polarization, in a) for $P_{e^-}=-80\%$ and in b)
for $P_{e^-}=+80\%$, as a function of positron
polarization, $P_{e^+}$ \ci{Ghodbane}. \la{fig_susy6}}}
\vspace{9cm}
\end{figure}

\begin{figure}
\hspace{-.8cm}
\begin{minipage}{7cm}
\begin{picture}(7,5)
\put(-.2,0){\includegraphics{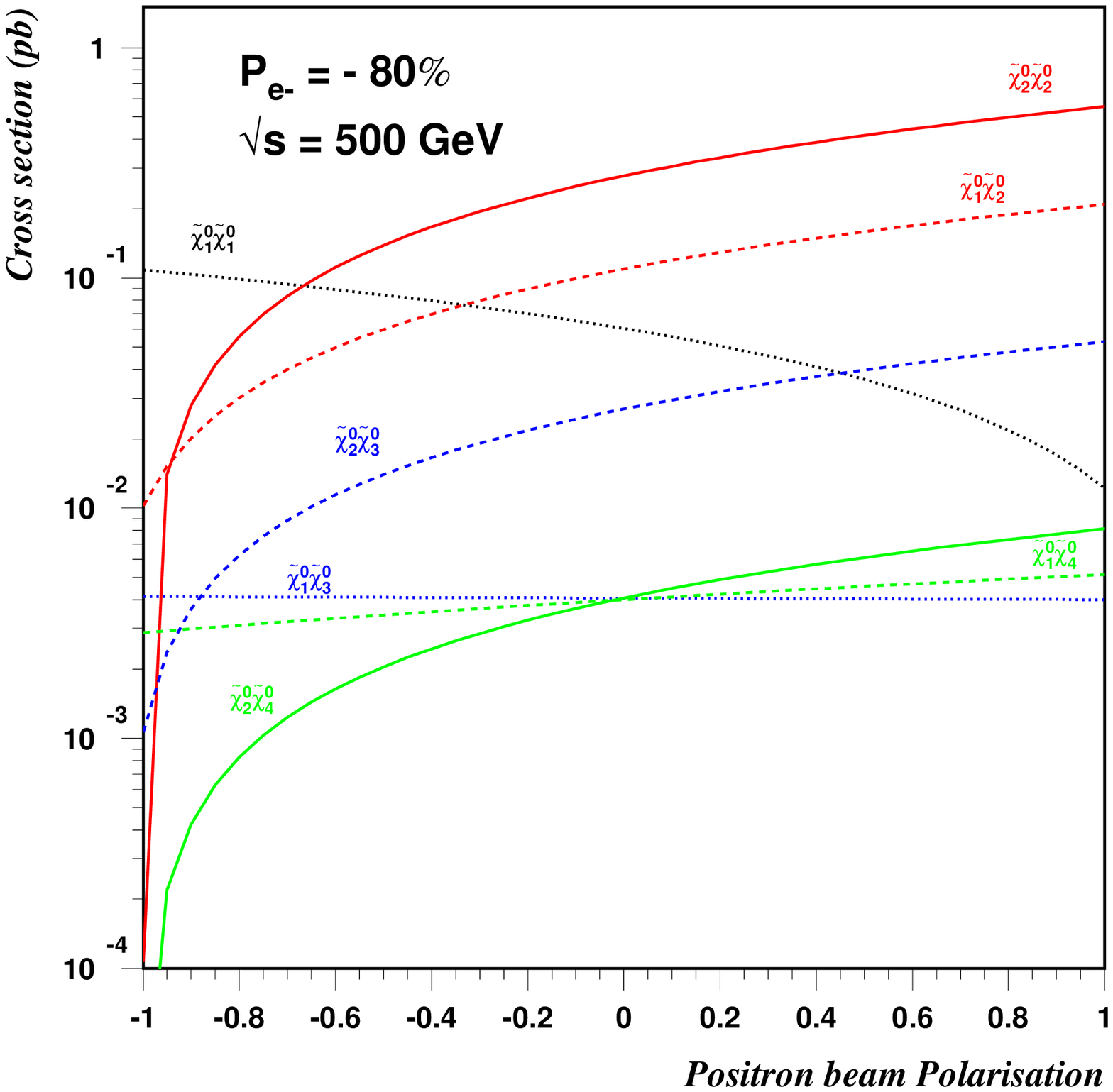}}
\put(4.5,4.5){\small a)}
\end{picture}\par\vspace{1cm}
\end{minipage}\hfill\hspace{.2cm}
\begin{minipage}{7cm}
\begin{picture}(7,5)
\put(-.2,0){\includegraphics{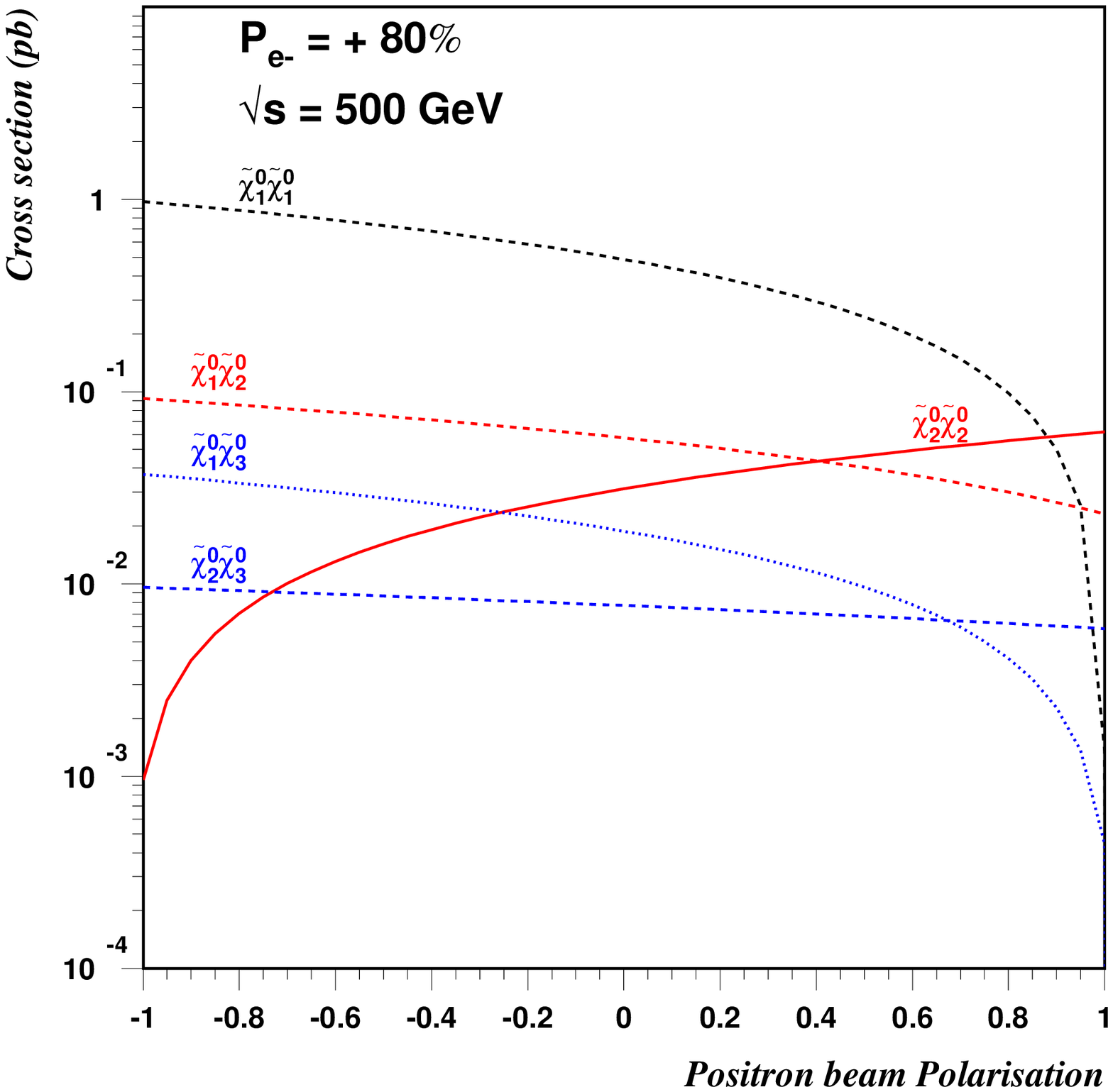}}
\put(3,4.5){\small b)}
\end{picture}\par\vspace{1cm}
\end{minipage}
\hspace*{-1cm}\parbox{14.7cm}{\caption{SUSY -- Neutralino production: 
Cross sections for 
$e^+ e^-\to \tilde{\chi}^0_{i}\tilde{\tilde{\chi}}^0_{j}$  [pb] at 
$\sqrt{s}=500$~GeV in our reference scenario are shown
  for fixed electron polarization, in a) for $P_{e^-}=-80\%$
and in b) for $P_{e^-}=+80\%$ , and variable positron
polarization, $P_{e^+}$ \ci{Ghodbane2}. \la{fig_susy2}}}\par\vspace*{-1cm}
%
\hspace{-.5cm}
\begin{minipage}{7cm}
\begin{picture}(7,7)
\put(0,0){\includegraphics{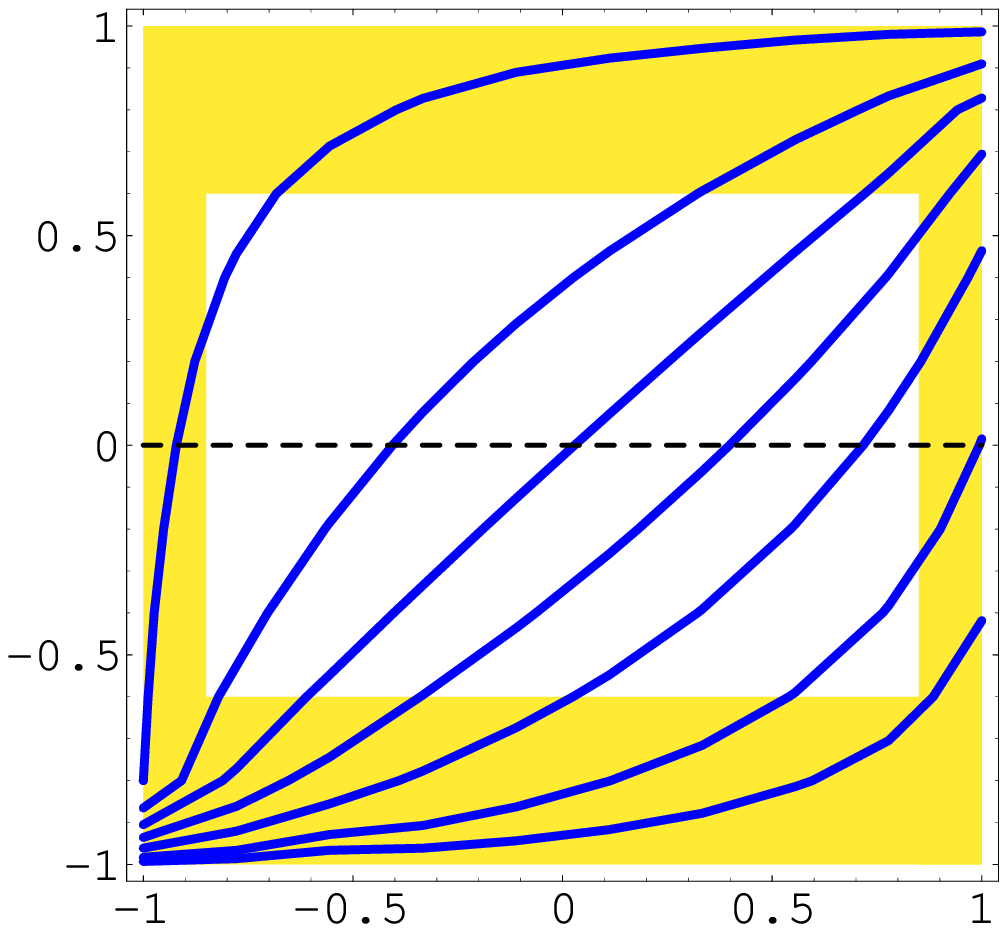}}
\put(.2,4.9){a) $m_{\tilde{e}_L}=176$~GeV, $m_{\tilde{e}_R}=132$~GeV}
\put(6,-1.2){\small $ P_{e^-}$}
\put(0,4.1){\small $ P_{e^+}$}
\put(3.8,1.6){\tiny 0}
\put(4.1,.3){\tiny 8\%}
\put(3.6,.8){\tiny 4\%}
\put(4.5,0){\tiny 10\%}
\put(3.,2.1){\tiny $-4\%$}
\put(2.5,2.7){\tiny $-8\%$}
\put(1.2,3.8){\tiny $-12\%$}
\end{picture}\par\vspace{.7cm}
\hspace{-.5cm}
\end{minipage}
\hspace*{-.5cm}
\begin{minipage}{7cm}
\begin{picture}(7,7)
\put(0,0){\includegraphics{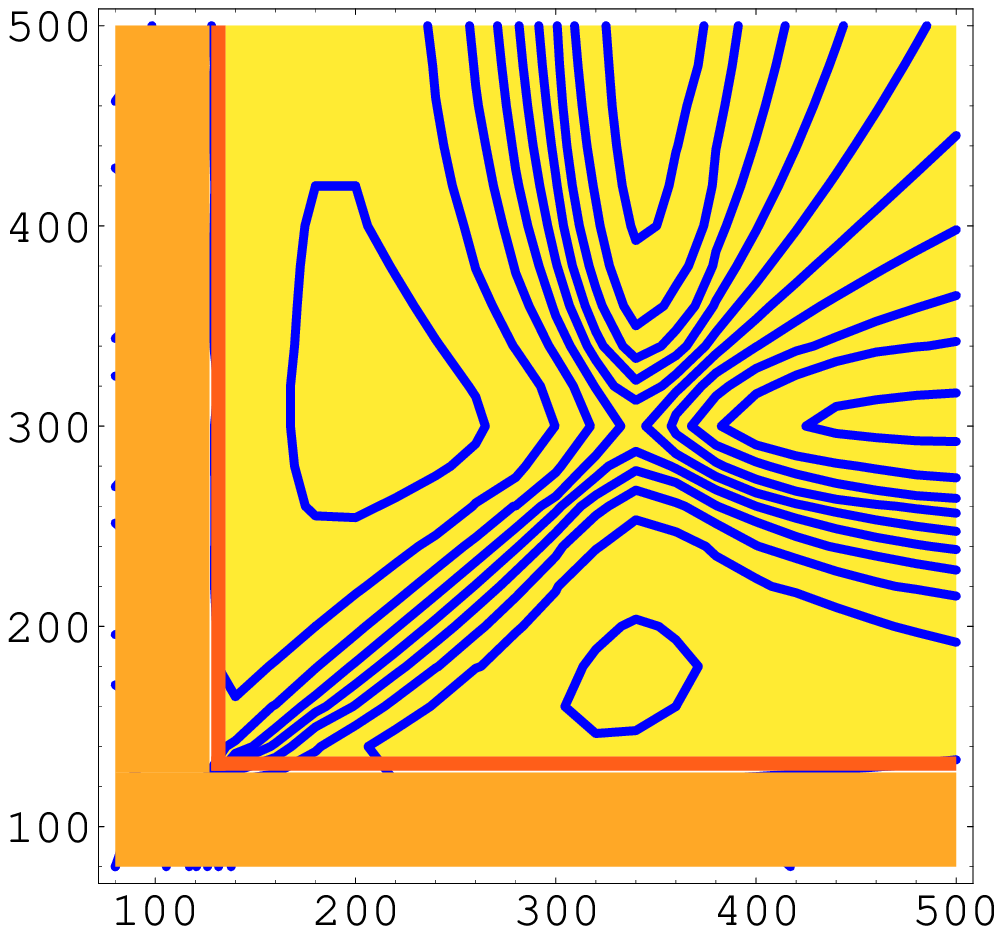}}
\put(.6,4.9){b) $P_{e^-}=-85\%$, $P_{e^+}=+60\%$}
\put(5.5,-1.2){\small $ m_{\tilde{e}_L}$}
\put(-.3,4.2){\small $ m_{\tilde{e}_R}$}
\put(2.2,2.3){\tiny $19$}
\put(2.1,4.2){\tiny $15$}
\put(3.7,4.35){\tiny $-20$}
\put(5.1,4.35){\tiny $-5$}
\put(5.4,3.9){\tiny 0}
\put(5.4,3.25){\tiny $5$}
\put(5.25,2.35){\tiny $19$}
\put(5.5,2.1){\tiny $23$}
\put(3.6,1){\tiny $-15$}
\put(3.5,.5){\tiny $-20$}
\end{picture}\par\vspace{1.2cm}
\end{minipage}
\hspace*{-1cm}\parbox{14.7cm}{\caption{ SUSY -- Determination of the 
selectron masses
$m_{\tilde{e}_{L,R}}$ in neutralino production and decay:
Contour lines of the forward--backward
asymmetry of the decay electron 
$A_{FB}$/\% of
$e^+ e^-\to\tilde{\chi}^0_1\tilde{\chi}^0_2, \tilde{\chi}^0_2\to 
\tilde{\chi}^0_1 e^+ e^-$
at $\sqrt{s}=(m_{\tilde{\chi}^0_1}+m_{\tilde{\chi}^0_2})+30$~GeV
in the reference scenario as a function of a) $P_{e^-}$ and $P_{e^+}$ for fixed 
$m_{\tilde{e}_L}=176$~GeV, $m_{\tilde{e}_R}=132$~GeV, 
and in b) $m_{\tilde{e}_L}$ and $m_{\tilde{e}_R}$ for fixed
$P_{e^-}=-85\%$, $P_{e^+}=+60\%$. Outside the red coloured region
direct production of $\tilde{e}_L$ or
$\tilde{e}_R$ is not possible, 
$\sqrt{s}/2<m_{\tilde{e}_{L,R}}$ \ci{Gudi_Char_Neut}.\la{fig_susy3}}}
\end{figure}

\begin{figure}[t]
\hspace{-1.1cm}
\begin{minipage}{7cm}
\begin{picture}(7,5)
\put(0,0){\includegraphics{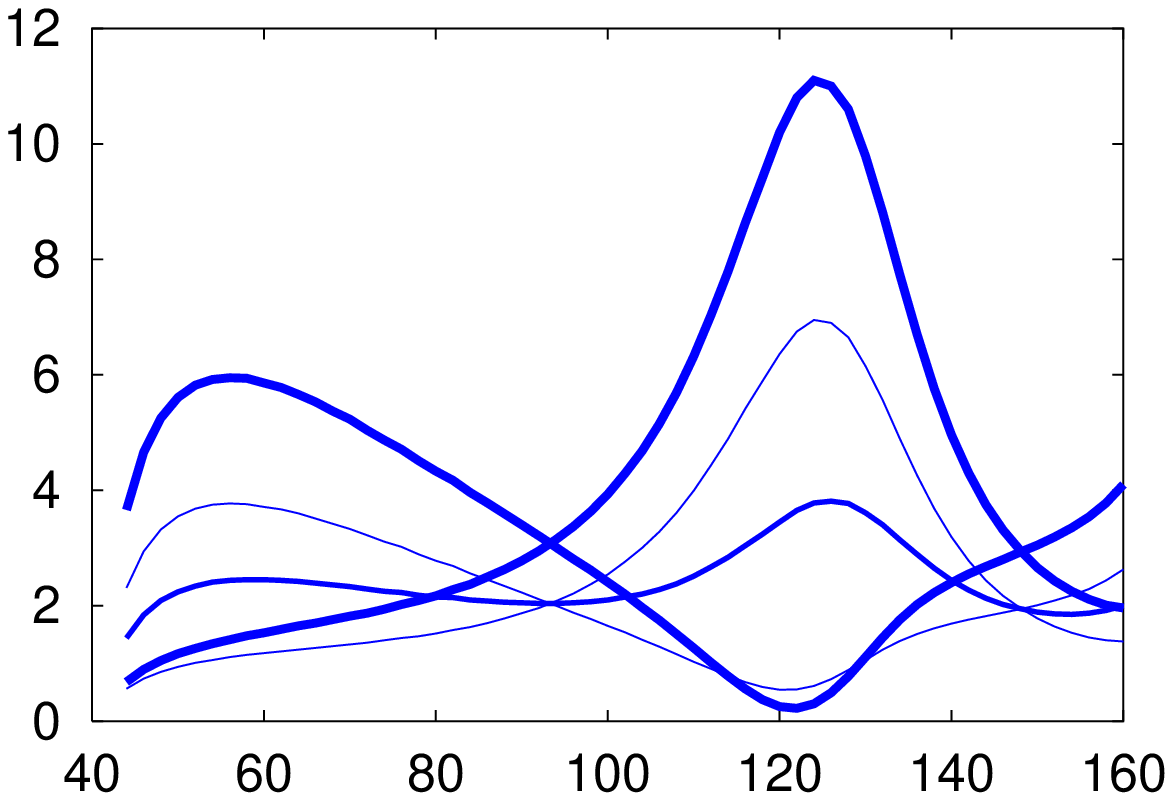}}
\put(5.8,-.3){ \small $M_1${\small /GeV}}
\put(.1,5.2){\small $ \sigma_e${\small /fb}}
\put(4.7,2.1){\small $(00)$}
\put(4.6,3.3){\small (R0)}
\put(5.5,4){\small (RL)}
\put(1.1,2.1){\small (L0)}
\put(1.1,2.9){\small (LR)}
\end{picture}\par\vspace{.3cm}
\end{minipage}\hfill\hspace{.2cm}
\begin{minipage}{7cm}
\begin{picture}(7,5)
\put(-.2,0){\includegraphics{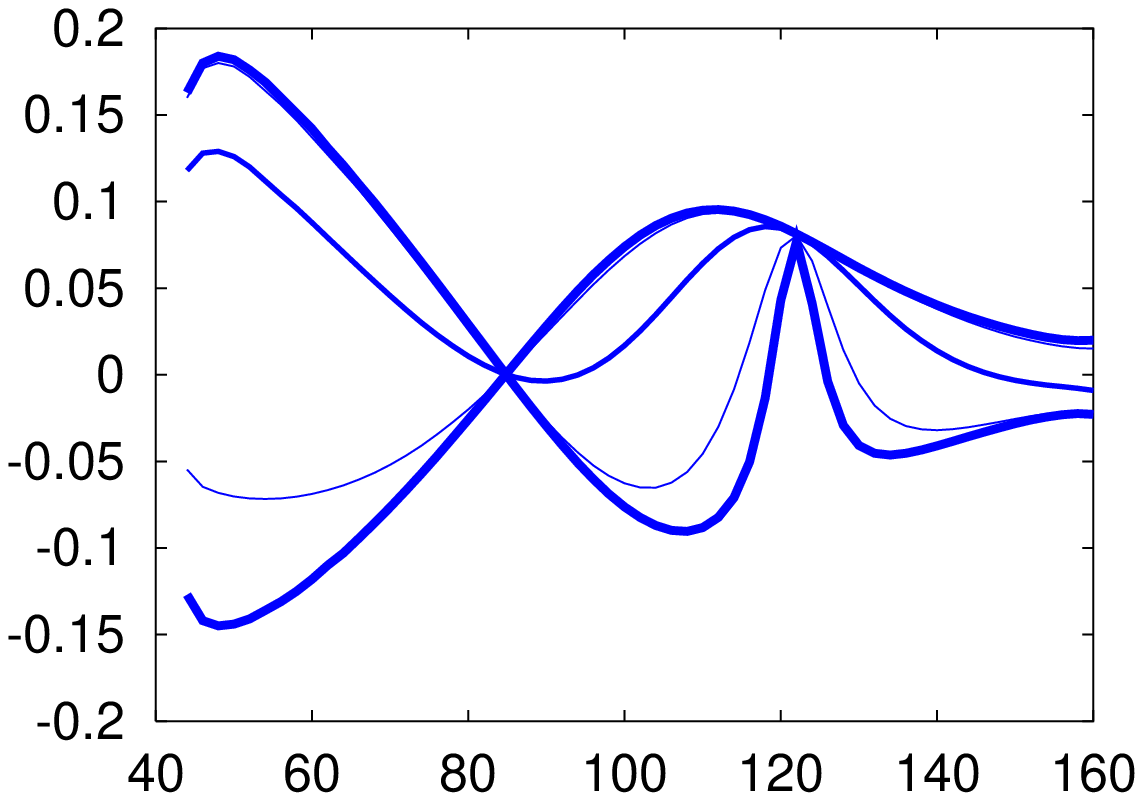}}
\put(5.5,-.3){ \small $ M_1${\small /GeV}}
\put(.2,5.2){\small $ A_{FB}$}
\put(1.9,4.4){\small (LR)$\approx (-0)$}
\put(4,1.4){\small (LR)}
\put(3.6,2.4){\small (L0)}
\put(4.5,4){\small (RL)$\approx$(R0)}
\put(1.1,3.5){\small $(00)$}
\put(1.2,2.2){\small (R0)}
\put(1.9,1){\small (RL)}
\end{picture}\par\vspace{.3cm}
\end{minipage}
\hspace*{-1cm}\parbox{14.7cm}{\caption{SUSY -- Determination of $M_1$ 
parameter in neutralino production and decay:
In a) cross sections $\sigma\times BR$ and in b)
the forward--backward asymmetries $A_{FB}$ of the decay electron
of $e^+ e^-\to \tilde{\chi}^0_1 \tilde{\chi}^0_2$, 
$\tilde{\chi}^0_2\to \tilde{\chi}^0_1 e^+ e^-$ are shown
at $\sqrt{s}=m_{\tilde{\chi}^0_1}+m_{\tilde{\chi}^0_2}+30$~GeV as function of 
gaugino parameter $M_1$ for unpolarized beams (00), 
for only electron beam polarized 
(L0), (R0) with $P_{e^-}=\pm 85\%$ and for both beams 
polarized (LR), (RL) 
with $P_-=\mp 85\%$, $P_{e^+}=\pm 60\%$ \ci{Gudi_Char_Neut}.  
Other SUSY parameters as in the reference scenario. \la{fig_susy4}}}
\end{figure}

\begin{figure}
\hspace{-.5cm}
\begin{minipage}{7cm}
\begin{picture}(7,5)
\put(-.2,0){\includegraphics{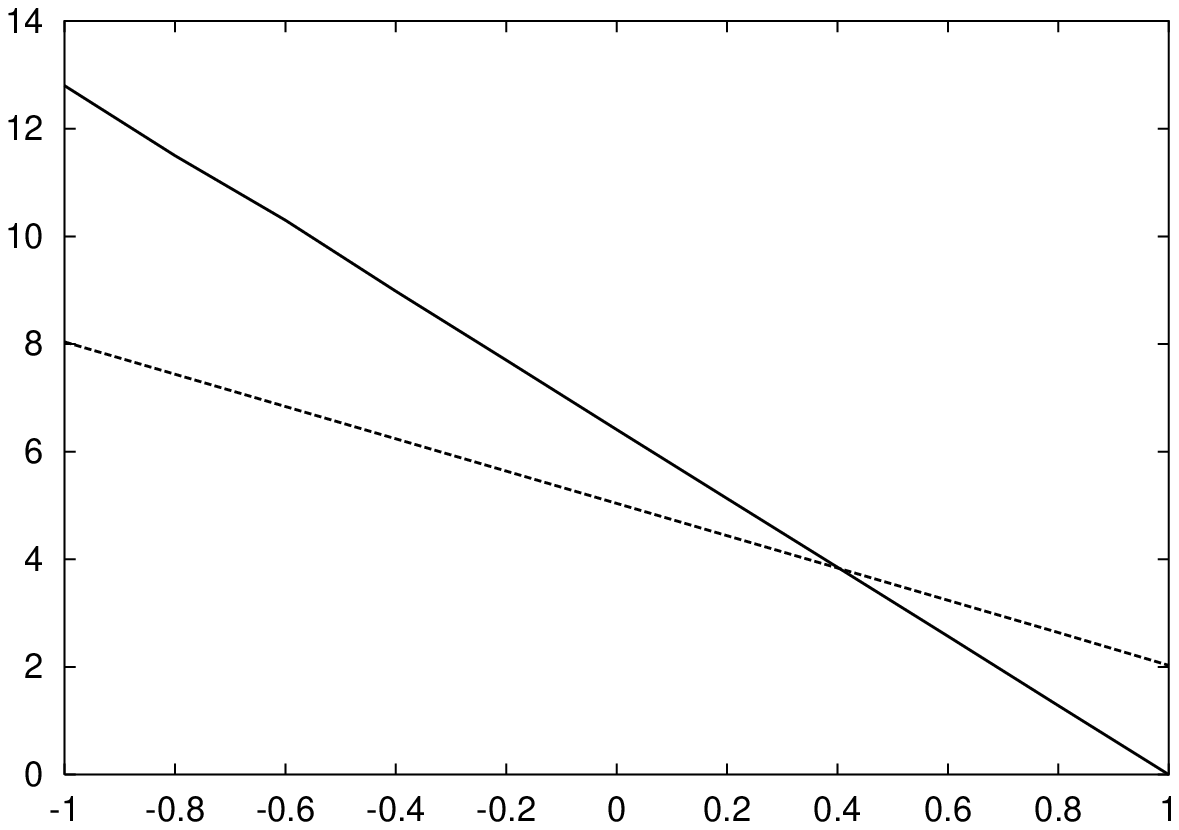}}
\put(4,3.7){\small $P_{e^-}=+80\%$}
\put(2.5,4.5){\small $e^+ e^-\to \tilde{\chi}^0_1 \tilde{\chi}^0_2$}
\put(1.5,1.5){ MSSM}
\put(1.6,3.5){NMSSM}
\put(5.6,-.3){\small $P_{e^+}$}
\put(-.1,4){\small $\sigma$} 
\put(-.3,3.5){\small [fb]}
\end{picture}\par\vspace{.2cm}
\end{minipage}\hfill\hspace{.8cm}
\begin{minipage}{7cm}
\begin{picture}(7,5)
\put(-.2,0){\includegraphics{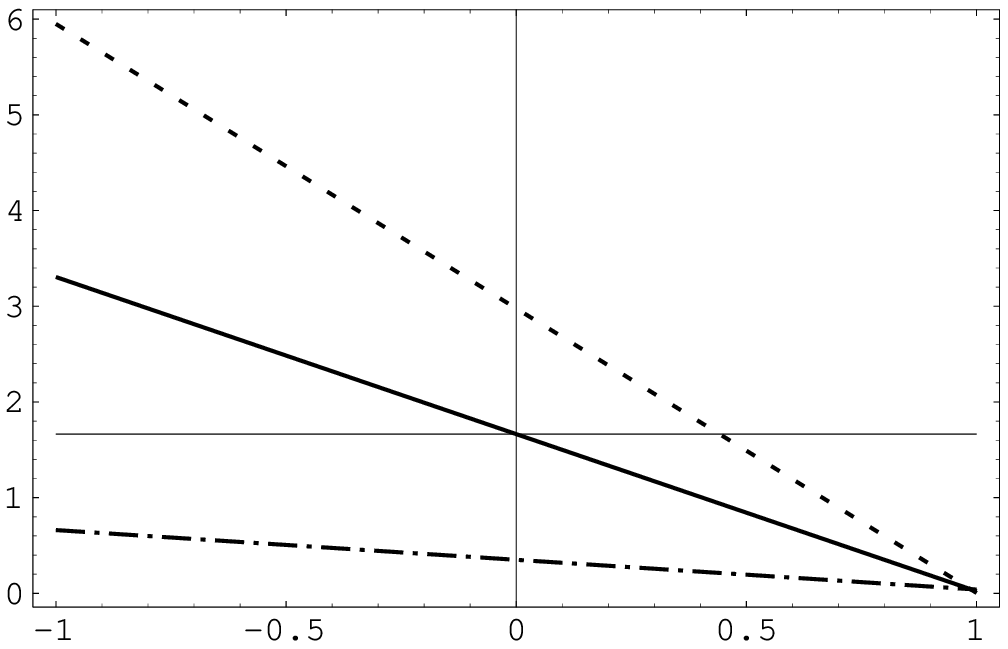}}
\put(-.7,4){\small $\sigma$}
\put(1.5,4.4){\small $e^+ e^-\to \tilde{\chi}^0_1 \tilde{\chi}^0_2$}
\put(-.8,3.5){\small [fb]}
\put(4.8,-.3){\small $P_{e^+}$}
\put(3,3.5){`E$_6$'--model}
\put(.2,.9){\small $P_{e^-}=-80\%$}
\put(.5,3.8){\small $P_{e^-}=+80\%$}
\put(.5,2.4){\small $P_{e^-}=0$}
\put(.1,1.5){\footnotesize unpolarized}
\end{picture}\par\vspace{.3cm}
\end{minipage}
\hspace*{-1cm}\parbox{14.7cm}{\caption{ SUSY -- Comparison of 
neutralino production in different SUSY models:
Cross sections $e^+ e^-\to \tilde{\chi}^0_1 \tilde{\chi}^0_2$ in fb
at $\sqrt{s}=500$~GeV in a) in the NMSSM and MSSM for fixed electron
polarization $P_{e-}=+80\%$ and in b) in the E$_6$--model for fixed
$P_{e^-}=0,\pm 80\%$ and variable positron polarization $P_{e^+}$
\ci{Franke}.  \la{fig_susy1}} }
\end{figure}

\end{document}